\documentclass[fleqn,usenatbib,useAMS]{mnras}

\usepackage{graphicx}	
\usepackage{amsmath}	
\usepackage{amssymb}	
\usepackage{multicol}        
\usepackage{bm}		
\usepackage{pdflscape}	

\usepackage{longtable}
\usepackage{epsf}
\usepackage{hyperref}
\hypersetup{colorlinks=true,linkcolor=blue,citecolor=blue,filecolor=blue,urlcolor=blue}
\usepackage{cleveref}
\usepackage{natbib}
\usepackage{chngcntr}

\newcommand{\Msun}{{\rm M}_{\odot}}
\newcommand{\Zsun}{{\rm Z}_{\odot}}

\newcommand{\tage}{\tau_{\rm age}}
\newcommand{\pgal}{\theta_{\rm gal}}
\newcommand{\pmah}{\theta_{\rm MAH}}
\newcommand{\psfh}{\theta_{\rm SFH}}
\newcommand{\pms}{\theta_{\rm ms}}
\newcommand{\pq}{\theta_{\rm q}}
\newcommand{\pmet}{\theta_{\rm Z}}
\newcommand{\pneb}{\theta_{\rm neb}}
\newcommand{\pdust}{\theta_{\rm dust}}
\newcommand{\zobs}{z_{\rm obs}}

\newcommand{\dd}{{\rm d}}
\newcommand{\Fatt}{F_{\rm att}}

\newcommand{\probsfh}{P_{\rm SFH}}
\newcommand{\probmet}{P_{\rm met}}
\newcommand{\probneb}{P_{\rm neb}}

\newcommand{\fsurv}{F_{\rm surv}}
\newcommand{\frem}{F_{\rm rem}}
\newcommand{\lssp}{L_{\rm SSP}}
\newcommand{\lcsp}{L_{\rm CSP}}
\newcommand{\lss}{L_{\rm SS}}

\newcommand{\imf}{\Phi_{\rm IMF}}

\newcommand{\us}{u_{\rm s}}
\newcommand{\Us}{U_{\rm s}}

\newcommand{\beq}{\begin{eqnarray}}
\newcommand{\eeq}{\end{eqnarray}}
\newcommand{\ben}{\begin{enumerate}}
\newcommand{\een}{\end{enumerate}}
\newcommand{\bit}{\begin{itemize}}
\newcommand{\eit}{\end{itemize}}

\newcommand{\Mhalo}{M_{\rm halo}}
\newcommand{\Mstar}{M_{\star}}

\newcommand{\Mstari}{M_{\star}^{i}}

\newcommand{\qdrop}{q_{\rm drop}}
\newcommand{\qtime}{t_{\rm q}}
\newcommand{\mumdf}{\mu_{\rm mdf}}
\newcommand{\sigmdf}{\sigma_{\rm mdf}}

\newcommand{\terf}{\mathcal{T}_{\rm erf}}

\newcommand{\Av}{A_{\rm V}}

\newcommand{\Alam}{A_{\lambda}}
\newcommand{\klam}{k_{\lambda}}
\newcommand{\klamcal}{k_{\lambda,{\rm Cal}}}
\newcommand{\klamleh}{k_{\lambda,{\rm L+02}}}

\newcommand{\sfr}{\dot{\Mstar}}

\usepackage[table]{xcolor} 
\definecolor{hpurple}{HTML}{7E16DF}

\usepackage{xspace}
\newcommand{\tng}{IllustrisTNG\xspace}
\newcommand{\um}{UniverseMachine\xspace}
\newcommand{\dstar}{Diffstar\xspace}

\usepackage[T1]{fontenc}
\usepackage{ae,aecompl}
\usepackage{newtxtext,newtxmath}

\title[DSPS: Differentiable Stellar Population Synthesis]{DSPS: Differentiable Stellar Population Synthesis}

\author[A.P. Hearin, et al.]{Andrew P. Hearin$^{1}$\thanks{Contact e-mail: \href{mailto:ahearin@anl.gov}{ahearin@anl.gov}}, Jon\'{a}s Chaves-Montero$^{1,2,3}$, Alex Alarcon$^1$, Matthew R. Becker$^1$, Andrew Benson$^4$
\\
$^{1}$ HEP Division, Argonne National Laboratory, 9700 South Cass Avenue, Lemont, IL 60439, USA\\
$^2$Donostia International Physics Centre, Paseo Manuel de Lardizabal 4, 20018 Donostia-San Sebastian, Spain\\
$^3$Institut de F\'isica d'Altes Energies, The Barcelona Institute of Science and Technology, Campus UAB, E-08193 Bellaterra (Barcelona), Spain\\
$^4$Carnegie Observatories, 813 Santa Barbara Street, Pasadena, CA 91101
}
\date{}

\begin{document}
\label{firstpage}
\pagerange{\pageref{firstpage}--\pageref{lastpage}}
\maketitle

\begin{abstract}
Models of stellar population synthesis (SPS) are the fundamental tool that relates the physical properties of a galaxy to its spectral energy distribution (SED). In this paper, we present DSPS: a python package for stellar population synthesis. All of the functionality in DSPS is implemented natively in the JAX library for automatic differentiation, and so our predictions for galaxy photometry are fully differentiable, and directly inherit the performance benefits of JAX, including portability onto GPUs. DSPS also implements several novel features, such as {\it i)} a flexible empirical model for stellar metallicity that incorporates correlations with stellar age, {\it ii)} support for the \dstar model that provides a physically-motivated connection between the star formation history of a galaxy (SFH) and the mass assembly of its underlying dark matter halo. We detail a set of theoretical techniques for using autodiff to calculate gradients of predictions for galaxy SEDs with respect to SPS parameters that control a range of physical effects, including SFH, stellar metallicity, nebular emission, and dust attenuation. When forward modeling the colors of a synthetic galaxy population, we find that DSPS can provide a factor of 5 speedup over standard SPS codes on a CPU, and a factor of 300-400 on a modern GPU. When coupled with gradient-based techniques for optimization and inference, DSPS makes it practical to conduct expansive likelihood analyses of simulation-based models of the galaxy--halo connection that fully forward model galaxy spectra and photometry.
\end{abstract}

\section{Introduction}
\label{sec:intro}

Stellar population synthesis (SPS) is the prevailing framework for predicting the spectral energy distribution (SED) of a galaxy from its fundamental physical properties \citep{Conroy13}. SPS is a mature subfield with a long history \citep[e.g.,][]{Tinsley1978_sps, Bruzual1983_sps, Arimoto1987_sps, Buzzoni1989_sps,bruzual_charlot_1993,worthey_1994_age_metal_degeneracy,maraston_1998,Leitherer_starburst99}, and applications of SPS range from inferring the physical properties of individual galaxies \citep{sawicki_yee_1998,brinchmann_ellis_2000,salim_etal07_sdss_sfrs,kriek_etal09,leja_etal19_3dhst}, to forward modeling the galaxy distribution across cosmic time \citep{baugh_cole_frenk_1996,kauffmann_etal99,somerville_primack99,korytov_etal19,drakos_etal21}. In reflection of the scientific breadth of this topic, there are by now many publicly available libraries that can be used to carry out SPS-related calculations \citep[e.g.,][]{Bruzual2003,fioc_rocca_volmerange99_pegase,leborgne_etal04,maraston05,conroy_gunn_white_2009,eldridge_etal17,johnson_etal21_prospector,larry_bradley_2020_photutils_v1p0}.

A wide range of theoretical models and techniques are considered to be components of SPS, from the specialized interpolation algorithms used to transform the outputs of stellar evolution codes into isochrones and stellar evolution tracks \citep{morton_2015_isochrones,dotter16_mesa_isochrones}, to the modeling of a galaxy's chemical evolution history \citep{audouze_tinsley_1976, weinberg_etal17}, to the computation of the observed photometry of a galaxy from its composite stellar population \citep{oke_sandage_1968,blanton_etal97_kcorrect}. These and other related predictions of SPS models can be computationally intensive, which can create a significant limitation on the level of physical realism that is achievable in a practical analysis. Classical machine learning (ML) techniques such as Gaussian Process emulation \citep{rasmussen_williams_2006_gp_review,gpy2014,george_gp_citation} or polynomial chaos expansion \citep{wiener_1938,xiu_2010_pce_review} are well-suited to address these computational challenges. The basic idea underlying this application of ML is to carry out a pre-processing step in which a limited number of expensive function evaluations are used to optimize the hyper-parameters of the ML model; subsequently, one uses the ML model to provide a fast-evaluating, ``surrogate" prediction in the performance-critical bottleneck of the analysis (hereafter, we will use the term ``surrogate function" to refer in a generic sense to an efficient ML-based approximation to some other function). Such ML ``emulation" methods have been used within the field of cosmology for years \citep{heitmann_etal06_cosmic_calibration}, and are now used by many different groups to accelerate cosmological predictions in a wide variety of contexts \citep[e.g.,][]{harnois_deraps_2019,mcclintock_etal19,euclid_emulator_2019,nishimichi_2019_dark_quest,kokron_etal21,ramachandra_etal21}. Recently, artificial intelligence algorithms (AI) such as a neural network \citep{mcculloch_pitts_1943_neural_network} have become more widely deployed as surrogate functions in cosmology and astrophysics \citep[e.g.,][]{kobayashi_etal20,arico_etal21_bacco,villaescusa_navarro_etal20_camels1}, including applications to stellar population synthesis \citep[][]{alsing_etal20_speculator}. 

When making a computationally expensive prediction, using an AI/ML-based surrogate function not only reduces the runtime of any particular model evaluation, but has the additional benefit of being an exactly {\it differentiable} function of its parameters, a highly advantageous feature in likelihood analyses. Gradient-based optimization algorithms such as BFGS \citep{broyden_1970_B_in_BFGS,fletcher_1970_F_in_BFGS,goldfarb_1970_G_in_BFGS,shanno_1970_S_in_BFGS} and Adam \citep{kingma_ba_adam_2015} require far fewer total evaluations of the objective function relative to other techniques. Inference algorithms such as Hamiltonian Monte Carlo \citep[HMC,][]{duane_etal87_hmc} offer analogous reductions in the number of likelihood evaluations required to derive posteriors relative to conventional MCMC methods. These performance benefits become increasingly dramatic in analyses of models with larger numbers of free parameters. For example, contemporary optimization algorithms leverage the differentiability of a neural network to train its weights and biases, which can comprise millions of degrees of freedom, and HMC is routinely used to derive posteriors on models with hundreds of parameters or more \citep{hoffman_gelman_2014_nuts}.

As an alternative to emulating the predictions of a model, the benefits of differentiability can also be reaped by directly implementing the model in a software library that supports automatic differentiation. Autodiff is an algorithm for evaluating the derivative of a function defined by a computer program, and is a distinct algorithm from numerically estimating a derivative via finite differencing methods \citep[see][and references therein]{baydin_etal15}. The computational cost of finite differencing estimates scales with the number of parameters, which is not the case for autodiff algorithms, and so gradient computations with autodiff vastly outperform finite-difference derivatives as the dimension of the model parameter space increases. Whereas numerical derivatives require careful checking of the finite step size to protect against inaccuracy (which can be especially tedious for the case of higher-order derivatives), the autodiff algorithm ensures that the gradient computation gives the same result as symbolic differentiation within working precision. 

There are numerous publicly available libraries providing high-performance implementations, such as TensorFlow \citep{tensorflow2015_whitepaper}, PyTorch \citep{pytorch_citation_2019}, and JAX \citep{jax2018github}; these libraries provide a convenient Python interface to autodiff algorithms that are highly performant on both CPUs and GPUs. Autodiff computations have now been used in a wide range of applications in scientific computing, including cosmological N-body simulations \citep{modi_lanusse_seljak_2021_flowpm}, molecular dynamics \citep{schoenholz_cubuk_2019_jaxmd}, and fluid dynamics \citep{kochkov_etal21}.

Motivated by these developments, in this paper we present DSPS, a stellar population synthesis code implemented in JAX. Although code written in JAX is very much like standard Numpy-based python, JAX has been specifically designed to target GPUs and other accelerator devices, and so there are numerous kinds of computations that require a specialized implementation (we refer the reader to \href{https://jax.readthedocs.io/en/latest/}{the JAX documentation} for further details). For this reason, implementing standardized SPS computations in JAX is a non-trivial effort, and throughout the paper we will use the terminology that a function has a ``differentiable implementation" to mean that the operations required to evaluate the function can be formulated within the restrictions of a library for automatic differentiation such as JAX; we will similarly describe a computation as being ``differentiable" when it has been carried out based on autodiff. Thus one of the primary purposes of DSPS is to make publicly available our JAX-based reimplementation of many of the standard computations in SPS. In particular, most of the models and calculations in DSPS have become widely standardized and were developed long ago in pioneering works of SPS such as \citet{Tinsley1978_sps,bruzual_charlot_1993,worthey_1994_age_metal_degeneracy}. As we will discuss throughout the paper, there are numerous benefits to our JAX-based SPS calculations, including GPU-accelerated performance, good scaling with the dimension of the model parameter space, and the simplification of analyses utilizing gradient information. 

This paper is organized as follows. In \S\ref{sec:sps} we describe how SPS predictions for galaxy SEDs are natural to implement with autodiff, and in \S\ref{sec:ingredients}, we illustrate autodiff-based techniques for calculating gradients of galaxy SEDs with respect to various physical ingredients that have differentiable implementations in DSPS. We discuss the advantages and limitations of DSPS in \S\ref{sec:discussion}, and conclude with a brief summary in \S\ref{sec:summary}.

\section{Differentiable Stellar Population Synthesis}
\label{sec:sps}

In this section, we describe our approach for making differentiable predictions for the SED of a galaxy based on stellar population synthesis. Here we will focus on general considerations of autodiff-based SPS calculations; in \S\ref{sec:ingredients} we will provide numerous examples of predictions for SEDs that are differentiable\footnote{We remind the reader that here and throughout the paper, we use the term ``differentiable" to mean ``implemented in an autodiff library".} with respect to specific modeling ingredients.

In a forward modeling analysis of the SED of some observed galaxy, $L_{\rm obs}(\lambda),$ one varies model parameters $\pgal$ and generates predictions $L_{\rm pred}(\lambda\vert\pgal).$ When estimating confidence intervals on $\pgal,$ the calculation of {\it gradients}, $\partial L(\lambda)/\partial\pgal,$ is not required by conventional MCMC methods, which need only evaluate the likelihood function itself, not its derivatives. But as discussed in \S\ref{sec:intro}, advances in Bayesian inference such as Hamiltonian Monte Carlo can dramatically outperform traditional MCMC when such gradients are available, and the same is true of optimization analyses in which one is primarily interested in estimating the best-fit point in parameter space. In all contemporary SPS codes, these gradients must be approximated numerically using finite-differencing methods. In \S\ref{subsec:diffseds} we describe how the SPS framework for predicting galaxy SEDs admits a natural implementation in autodiff, enabling calculations of $\partial L(\lambda)/\partial\pgal$ that are exact, efficient, and scalable onto high-performance computing resources. In \S\ref{subsec:photometry}, we show that predictions for SED-derived quantities such as photometry or emission line strength can also be differentiably formulated. 

Here and throughout the paper, we will use the variable $\tage$ to denote the length of time that has passed since the birth of a star or stellar population. The age of the universe is the physically natural time variable used to describe the star formation history of a galaxy, and so we will use the variable $t$ to refer to the age of the universe at the time of some particular event.

\subsection{Differentiable Galaxy SEDs}
\label{subsec:diffseds}

One of the foundational concepts of stellar population synthesis is the Simple Stellar Population (SSP), which is defined to be a population of stars that formed simultaneously from a homogeneous gas cloud; by definition, all the stars in an SSP have the same age and metallicity. If we define $\lss(\lambda\vert\Mstari, \tage, Z)$ in units of $L_{\odot}{\rm Hz}^{-1}$ to be the luminosity per unit frequency emitted at time $\tage$ after the birth of a single star with initial mass $\Mstari,$ and initial metallicity $Z,$ then we have
\beq
\label{eq:sspdef}
\lssp(\lambda\vert \tage,Z)\equiv\int\dd\Mstari\imf(\Mstari)\lss(\lambda\vert\Mstari,\tage,Z),
\eeq
where $\imf$ is the number of stars per unit mass in the SSP, i.e., the initial mass function (IMF); Eq.~\ref{eq:sspdef} is normalized per unit of total stellar mass formed.\footnote{For the sake of brevity, in this section and throughout the paper we assume that the initial mass, $\Mstari,$ and total metallicity, $Z,$ uniquely determine the SED emitted by a star at a time $\tage$ after it is born, when in fact numerous other variables are known to play a significant role, e.g., the rotation speed and chemical abundance pattern. We furthermore neglect treatment of the effect of stellar binaries on the light emitted by an SSP. See \S\ref{subsec:limits} for discussion of how DSPS could be extended to take these effects into account in a differentiable fashion.}

Deriving the left-hand side of Eq.~\ref{eq:sspdef} is one of the core computations of stellar population synthesis, and the calculation is rather involved. First, stellar evolution tracks must be computed using a stellar evolution code such as MESA \citep{paxton_etal11_mesa,paxton_etal13_mesa,paxton_etal15_mesa}. Due to the computational expense of stellar evolution calculations, these stellar tracks must be computed on a grid in advance and stored to disk. Calculating $\lss(\lambda\vert\Mstari,\tage,Z)$ from such stellar tracks is a highly nontrivial task due to the huge dynamic range in timescales spanned by stellar evolutionary physics, and so specialized interpolation techniques are required in order to calculate isochrone libraries such as MIST \citep{choi_etal16_mist} from the fundamental outputs of MESA \citep[see][and references therein]{dotter16_mesa_isochrones}. An isochrone table together with a library of stellar spectra then permits the calculation of $\lssp(\lambda\vert \tage,Z)$ via the IMF-weighted sum of $\lss(\lambda\vert\Mstari,\tage, Z)$ shown in Eq.~\ref{eq:sspdef}. The complex nature of this computation is one of the chief technical reasons that drives most SPS codes to make a fixed choice for the IMF, and to predict galaxy SEDs as weighted sums of a discrete collection of $\lssp(\lambda\vert \tage, Z)$ that are tabulated on a grid in advance.

Due to these considerations, in stellar population synthesis the SED of the composite stellar population of a galaxy, $\lcsp(\lambda),$ is calculated according to the following weighted sum:
\beq
\label{eq:cspdef}
\lcsp(\lambda)= \sum_{i}\lssp(\lambda\vert x_i)\cdot P(x_i),
\eeq
where $x_i$ is the finite (n-dimensional) grid of stellar population properties used in the pre-computed tabulation of $\lssp(\lambda),$ and $P(x_i)$ quantifies the fractional abundance of the SSP with properties $x_i$ within the composite population. For example, if age and metallicity are the only SSP properties in consideration, then the weighted sum over $x_i$ in Eq.~\ref{eq:cspdef} will be performed over a two-dimensional grid of $\tage$ and $Z,$ and $P(x_i)$ quantifies the abundance of stars in the galaxy as a function of age and metallicity. Or if one additionally incorporates the dependence of $\lssp(\lambda)$ upon the ionization state of the nebular gas, $\Us,$ then the finite summation in Eq.~\ref{eq:cspdef} will be carried out over a three-dimensional grid (as in \S\ref{subsec:nebulae}). 

Regardless of the details of the grid used to define the SSPs, in order to understand how to make differentiable predictions for the SED of a galaxy, the salient feature to focus on in Eq.~\ref{eq:cspdef} is that the collection of $\lssp(\lambda)$ are precomputed in advance and thereafter held fixed. Thus for a model of the SED of a galaxy with parameters $\pgal,$ any changes to model predictions for the SED are exclusively driven by how $\pgal$ modifies the {\it weights} in the summation over SSP spectra:
\beq
\label{eq:cspdeftheta}
\lcsp(\lambda\vert\pgal)= \sum_{i}\lssp(\lambda\vert x_i)\cdot P(x_i\vert\pgal).
\eeq
For example, model parameters $\pgal$ that modify the star formation history of the galaxy will change the distribution of stellar ages in the composite population, and thereby modify the $\tage$-dependence of $P(x_i\vert\pgal);$ in single-metallicity models one might directly have $Z$ as a model parameter $\pgal,$ or in more complex models $\pgal$ could regulate physical processes that influence chemical enrichment. 

From Eq.~\ref{eq:cspdeftheta} it is clear that gradients of the galaxy SED can be calculated exactly provided that one can calculate $\partial P(x_i)/\partial\pgal.$ One of the principal tasks handled by the DSPS library is providing differentiable calculations for the weights $P(x_i);$ as we will show throughout this paper, it is natural to achieve this differentiability in a wide variety of SPS calculations, provided that parametric models for $P(x_i)$ are suitably formulated and implemented. 

\subsection{Differentiable Predictions for SED-Derived Quantities}
\label{subsec:photometry}
In the previous section, we showed how calculation of the full SED of a galaxy can be formulated to admit a differentiable implementation. But galaxy samples targeted by cosmological surveys are commonly defined in terms of observable quantities that {\it derive} from the full SED, such as the photometric flux observed through a broad-band filter, or the strength of a particular emission line. We now describe differentiable techniques for predicting these two quantities in turn. 

\subsubsection{Photometry}
\label{subsubsec:photometry}

If the rest-frame SED of a galaxy is described in units of luminosity per unit frequency by $L_{\nu}(\lambda),$ and if we use $T_{\rm Q}(\lambda)$ to denote the filter transmission curve defined as the probability of photon transmission, then in the AB magnitude system, the absolute magnitude of the galaxy observed in the rest-frame through the filter, $M_{\rm Q},$ is given by
\beq
\label{eq:restmag}
M_{\rm Q} &=& -2.5\log_{10}\left(\frac{\int\frac{\dd\lambda}{\lambda}T_{\rm Q}(\lambda)L_{\nu}(\lambda)}{AB_0\int\frac{\dd\lambda}{\lambda}T_{\rm Q}(\lambda)}\right),
\eeq
where $AB_0=1.13492\times10^{-13} {\rm L}_{\odot}{\rm Hz^{-1}}$ is the the AB flux zero-point for a source observed at 10 pc \citep[see, e.g.,][for details]{oke_sandage_1968,hogg_etal02_kcorrection,blanton_etal97_kcorrect}. In Eq.~\ref{eq:restmag}, the integral is carried out across {\it wavelength}, but the integrand is $L_{\nu}(\lambda),$ the luminosity per unit {\it frequency}. We formulate the equations throughout the paper based on this convention, primarily because this is the form of the default SSP templates that are supplied by FSPS. Eq.~\ref{eq:restmag} can be directly compared to other common forms of this equation \citep[e.g., Equation 5 of][]{hogg_etal02_kcorrection} via the relations $\nu L_{\nu}=\lambda L_{\lambda}$ and ${\rm d}\lambda/\lambda=-{\rm d}\nu/\nu.$ We will henceforth drop the subscript from our notation, but it should be understood that $L(\lambda)=L_{\nu}(\lambda)$ throughout the paper.

Eq.~\ref{eq:restmag} is natural to implement in an autodiff library because the integral can be discretized and evaluated with trapezoidal summation, and so autodiff-based computations of rest-frame photometry are straightforward once one is equipped with a differentiable calculation of $L(\lambda).$ We remind the reader that in this section we restrict attention to general considerations regarding the differentiable calculation of SED-derived quantities such as $M_{\rm Q};$ in \S\ref{sec:ingredients} we provide numerous examples of photometry gradients with respect to specific modeling ingredients.

In order to predict the apparent magnitude of the galaxy in the observer-frame, $m_{\rm Q},$ we must take into account both the redshift of the SED, as well as the dimming of the source across cosmological distances. For the former, we need only apply the redshift relation, $\lambda_{\rm obs}=(1+\zobs)\lambda_{\rm rest},$ before carrying out the integral in the numerator of Eq.~\ref{eq:restmag}; this is straightforward to compute in JAX with simple interpolation; for notational convenience, let us denote the result by $M'_{\rm Q}.$ We then have $$m_{\rm Q}=M'_{\rm Q} + D_{\rm mod}(\zobs) -2.5\log_{10}(1+\zobs),$$ where $D_{\rm mod}(z)$ is the distance modulus function that depends on cosmology. The DSPS library implements a JAX-based calculation of $D_{\rm mod}(z)$ for flat $w{\rm CDM}$ cosmological models. Differentiability of $D_{\rm mod}(z)$ with respect to redshift is required of any application in which the galaxy properties are jointly fit together with $z;$ likewise, differentiability with respect to cosmological parameters is required by hierarchical inference applications that jointly fit for the summary statistics of a galaxy population \citep[e.g.,][]{leistedt_etal22_hierarchical_sps,alsing_etal22_nz}. For more general cosmological models beyond flat $w{\rm CDM}$, one could use \href{https://github.com/DifferentiableUniverseInitiative/jax\_cosmo}{jax-cosmo} in tandem with DSPS. For SPS analyses carried out at fixed cosmology, one could alternatively use JAX to interpolate from a lookup table for the function $D_{\rm mod}(z)$ that can be computed in advance with a publicly available package such as Astropy \citep{astropy_2013,astropy_2018} or Colossus \citep{diemer_2018_colossus}.

\subsubsection{Emission Line Strength}
\label{subsubsec:eqwidth}

A second characteristic of the galaxy SED that is commonly used in the selection of galaxy samples in cosmological surveys is the strength of a particular emission line at wavelength $\lambda,$ commonly quantified in terms of the equivalent width, $W_{\lambda},$ defined as
\beq
\label{eq:eqw}
W_{\lambda} = \int_{\lambda-\Delta\lambda}^{\lambda+\Delta\lambda}\dd\lambda'\left[L({\lambda}')-L_0({\lambda}')\right]/L_0(\lambda'),
\eeq
where $L(\lambda)$ is the spectrum of the galaxy, $\Delta\lambda$ specifies the wavelength range bracketing the line, and $L_0(\lambda)$ is the continuum spectrum evaluated at the center of the line. Since the integral in Eq.~\ref{eq:eqw} poses no problem for autodiff, then the key to a differentiable calculation of $W_{\lambda}$ is the computation of the continuum, $L_0(\lambda).$

To calculate $L_0(\lambda),$ we fit the spectrum $L(\lambda)$ in the neighborhood of the line with a quadratic polynomial, masking the wavelength range containing the line. General purpose fitting algorithms can be challenging to formulate in a differentiable fashion, but polynomial fitting is purely a matter of linear algebra, and so can be implemented in autodiff library without complications. Thus if the SED of a galaxy depends in a differentiable fashion upon some model parameter, $\theta,$ so that $\partial L(\lambda)/\partial\theta$ can be calculated with autodiff, then it is straightforward to calculate $\partial L_0(\lambda)/\partial\theta,$ since polynomial fitting is merely a sequence of linear operations, and so it is in turn straightforward to calculate $\partial W_\lambda/\partial\theta$ by differentiating through the right-hand side of Eq.~\ref{eq:eqw}. 

\section{Physical Ingredients and their Gradients}
\label{sec:ingredients}

The key take-away from the previous section is that the SED of a galaxy can be written as a probability-weighted sum over a precomputed grid of spectra of simple stellar populations, and so the differentiable calculation of the SED rests on the ability to compute gradients with respect to the weights (see Eq.~\ref{eq:cspdeftheta}). The specific form of these probability weights varies widely depending on the application, and so in \S\ref{sec:sps} we focused on general considerations concerning the differentiable formulation of SPS calculations.

We now turn attention to providing specific demonstrations of the differentiability of a diverse variety of models that are commonly encountered in SPS. We will limit our discussion to the physical ingredients that are currently implemented in DSPS, highlighting the autodiff-based techniques we use to calculate gradients with respect to the parameters of these models. We present differentiable techniques for calculating the dependence of galaxy SED upon stellar age in \S\ref{subsec:ages}, stellar metallicity in \S\ref{subsec:metals}, nebular emission in \S\ref{subsec:nebulae}, and dust attenuation in \S\ref{subsec:attenuation}. 

\subsection{Star Formation Histories and Stellar Ages}
\label{subsec:ages}
The light emitted by the stars in a galaxy at a time $t$ is the cumulative contribution of all the evolved simple stellar populations that formed prior to $t.$ Using $\lcsp(\lambda\vert t)$ to denote the SED of the composite stellar population, we have
\beq
\label{eq:spsagedef}
\lcsp(\lambda\vert t)= \int_{0}^{t}\dd t' \int_{0}^{Z_{\rm max}}\dd Z\cdot\sfr(t', Z)\cdot\lssp(\lambda\vert\tage, Z),
\eeq
where $\tage\equiv t-t',$ and the star formation history of the galaxy is $\sfr(t', Z).$\footnote{Note that the form of Eq.~\ref{eq:spsagedef} assumes that the SSP spectra are normalized per unit mass of stars formed. See Appendix~\ref{sec:remnants} for DSPS calculations of the {\it surviving} stellar mass that accounts for mass lost due to passive evolution.}  Throughout the remainder of \S\ref{subsec:ages}, we will work under the assumption that all the stars in a galaxy have the same metallicity, and postpone treatment of galaxy populations with a diverse chemical composition until \S\ref{subsec:metals}. Under this assumption, Equation~\ref{eq:spsagedef} becomes
\beq
\label{eq:csp}
\lcsp(\lambda\vert t)= \int_{0}^{t}\dd t'\sfr(t')\cdot\lssp(\lambda\vert\tage).
\eeq

As discussed in \S\ref{sec:sps}, the SEDs of simple stellar populations, $\lssp(\lambda\vert\tage),$ are typically pre-computed on a discretized grid of values of $\tage^{i},$ so that the integral in Eq.~\ref{eq:csp} becomes a finite summation:
\beq
\label{eq:cspsum}
\lcsp(\lambda\vert t) = \Mstar(t)\cdot\sum_{i}\lssp(\lambda\vert\tage^{i})\cdot\probsfh(\tage^{i}\vert t),
\eeq
where in Eq.~\ref{eq:cspsum}, $\probsfh(\tage^{i}\vert t)$ is the fraction of the stars at time $t$ that have ages between the boundaries of the $i^{\rm th}$ age bin,
\beq
\\
\probsfh(\tage^{i}\vert t) = \frac{1}{\Mstar(t)}\int_{t_{\rm lo}^{i}}^{t_{\rm hi}^{i}}\dd t'\sfr(t')\nonumber,
\eeq
with $t_{\rm hi/lo}^{i}=t'\pm\Delta\tage^{\rm i}/2,$ and 
\beq
\label{eq:mstartot}
\Mstar(t)\equiv\int_0^{t}\dd t'\sfr(t').
\eeq

Equation~\ref{eq:cspsum} shows that the influence of SFH upon the SED of a composite stellar population is calculable via a probability-weighted sum of the spectra of SSPs. In the computations implemented in DSPS, the finite collection of $\lssp(\lambda\vert\tage^{i})$ are loaded into memory as a contiguous block of data in advance\footnote{While this computational choice is widely used elsewhere in other libraries \citep[e.g., as in BAGPIPES][]{carnall_etal18_bagpipes}, some libraries such as Prospector \citep{johnson_etal21_prospector} enable analyses that supply SSPs on-the-fly on an as-needed basis.} and thereafter held fixed, and so the dependence of $\lcsp(\lambda)$ upon star formation history is entirely contained in the SFH-dependence of $\probsfh(\tage^{i}\vert t).$ Using $\psfh$ to denote the parameter(s) encoding the behavior of the star formation history of a galaxy, we can see that calculating $\partial\lcsp/\partial\psfh$ poses no significant technical obstacle for an autodiff-based implementation: the integrations in Eqs.~\ref{eq:cspsum}-\ref{eq:mstartot} can be evaluated numerically as a discretized sum, and so gradients of the SED can be achieved provided that $\sfr(t)$ admits a differentiable implementation. 

\begin{figure}
\includegraphics[width=8cm]{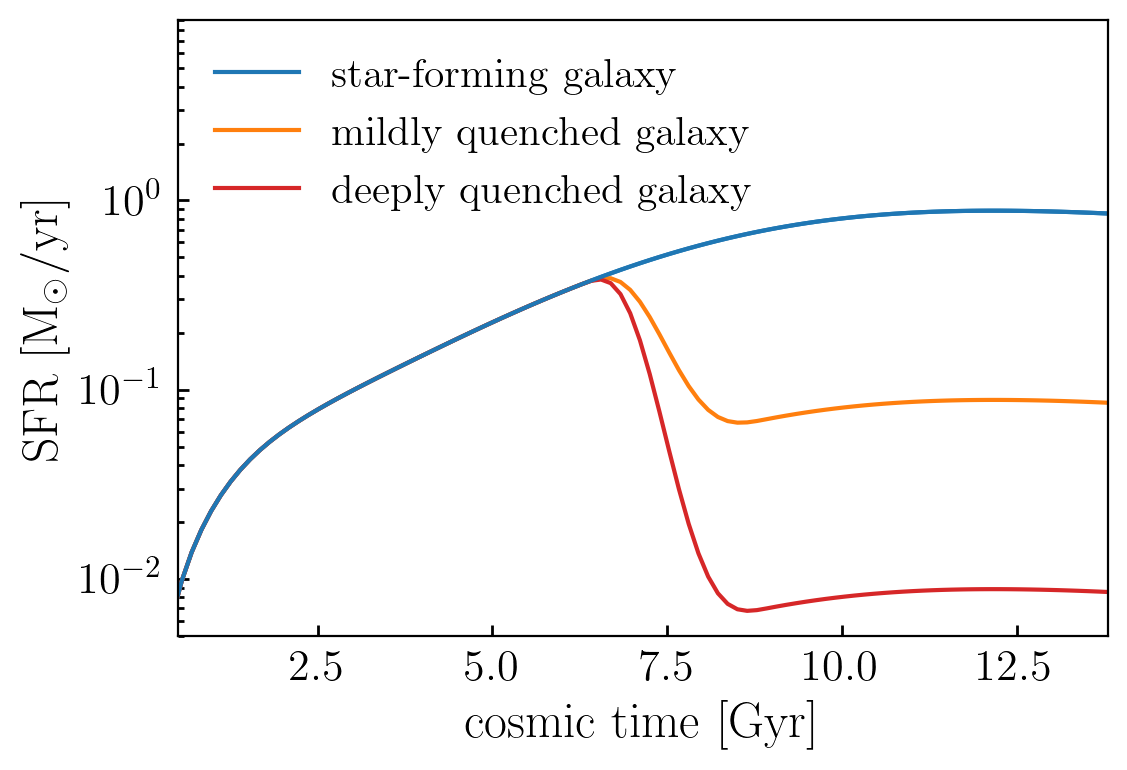}
\includegraphics[width=8cm]{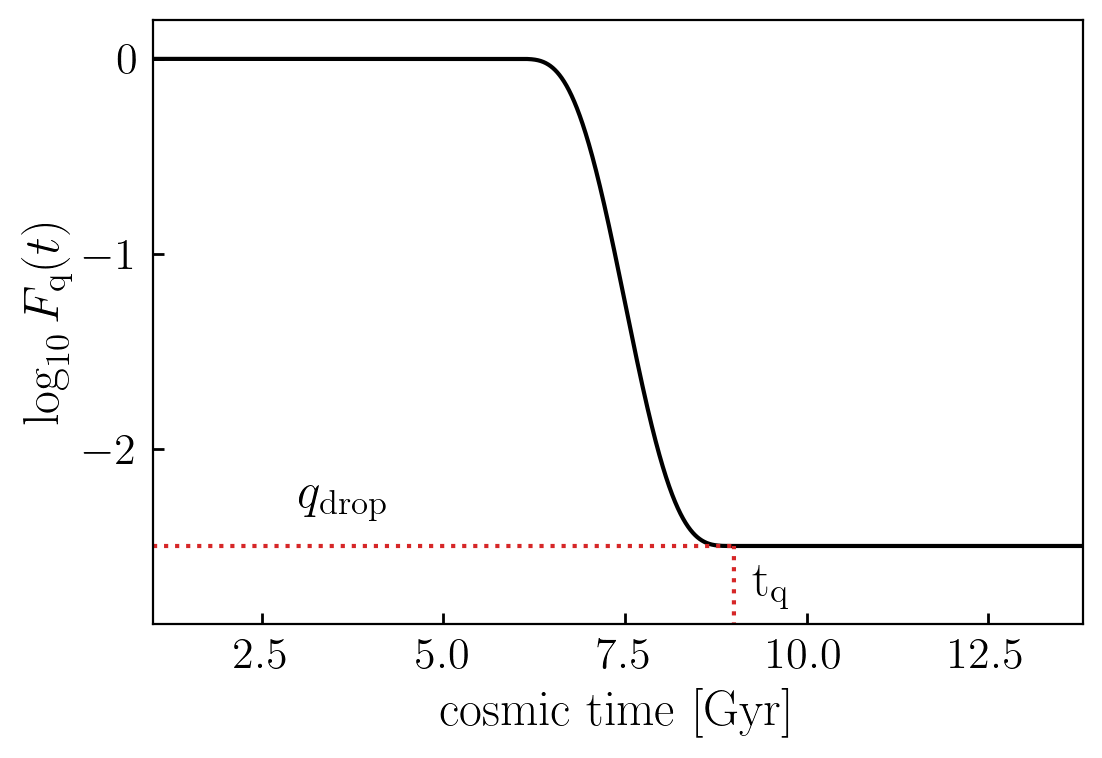}
\includegraphics[width=8cm]{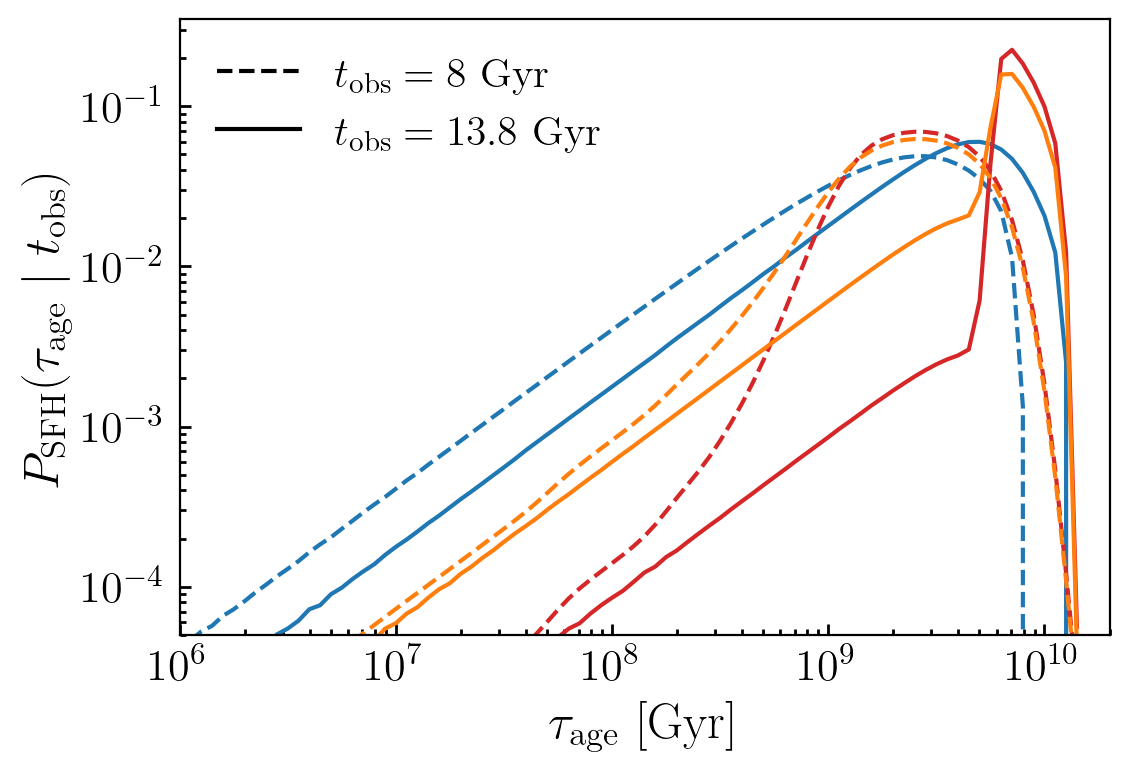}
\caption{{\bf Stellar populations of \dstar galaxies}. The top panel shows the SFH of three different galaxies parameterized by the \dstar model, each of which resides in a dark matter halo with $\Mhalo=10^{12}\Msun$ at redshift zero. The blue curve shows the history of a galaxy that remains on the main sequence for its entire lifetime; the red and orange curves show galaxies that experience a quenching event that shuts down star formation at $z\approx1.$ The middle panel illustrates the quenching function, $F_{\rm q}(t),$ used in the \dstar model to logarithmically drop the SFR of a quenched galaxy to some value, $\qdrop,$ below the main sequence (see Eq.~\ref{eq:diffstar}). The bottom panel compares the distribution of stellar ages of these same two galaxies observed at different redshifts as indicated in the legend.}
\label{fig:sfh_ages}
\end{figure}

All of the most widely-used parametric forms for SFH admit a straightforward implementation in JAX, including flexible, piecewise-defined models in which the SFH is calculated by interpolation between a set of control points in time. In order to demonstrate $\partial\lcsp/\partial\psfh,$ we will use the \dstar model for SFH introduced in \citet{alarcon_etal21}. The parameterization of \dstar was designed to have a close connection to the physics of galaxy formation, and to be sufficiently flexible to approximate the complex SFHs of galaxies predicted by contemporary simulations in an unbiased fashion. In the basic physical picture of the \dstar model, the accretion rate of gas, $\dot{M}_{\rm g},$ is proportional to the accretion rate of the dark matter halo, $\dot{M}_{\rm halo};$ main sequence galaxies transform accreted gas into stars over a gas consumption timescale, $\tau_{\rm cons},$ with efficiency, $\epsilon_{\rm ms};$ and some galaxies experience a quenching event that drops their SFR below the main-sequence rate by a multiplicative factor, $F_{\rm q}.$ The key equations that define the \dstar model are as follows:
\beq
\label{eq:diffstar}
&&\dot{M}_{\star}(t\vert\psfh) = \dot{M}_{\star}^{\rm ms}(t\vert\pms) \times F_{\rm q}(t\vert\pq)\nonumber\\ 
&&\dot{M}_{\star}^{\rm ms}(t\vert\pms,\pmah) = \epsilon_{\rm ms}(\pms)\int_{0}^{t}\dd t'\dot{M}_{\rm g}(t')F_{\rm cons}(t'\vert\tau_{\rm cons})\nonumber\\
&&\dot{M}_{\rm g}(t\vert\pmah) = f_{\rm b}\dot{M}_{\rm halo}(t\vert\pmah),
\eeq
where $f_{\rm b}=\Omega_{\rm b}/\Omega_{\rm m}$ is the cosmic baryon fraction, and $F_{\rm cons}(\tau_{\rm cons})$ controls how a fraction of the gas accreted onto the halo is gradually transformed into stars, and we have decomposed $\psfh$ into subspaces responsible for main-sequence evolution with $\pms,$ quenching with $\pq,$ and dark matter halo mass assembly history with $\pmah.$ For $\dot{M}_{\rm halo}(t),$ we use \href{https://github.com/ArgonneCPAC/diffmah}{\tt diffmah}, a differentiable parametric model that accurately captures the mass assembly of individual dark matter halos in both gravity-only and hydrodynamic simulations \citep[see][for details]{hearin_etal21_diffmah}. We refer the reader to \citet{alarcon_etal21} for a detailed description of the \dstar model.

In the top panel of Figure~\ref{fig:sfh_ages}, we show two different examples of SFHs in the \dstar model, each pertaining to a galaxy that resides in a dark matter halo with Milky Way mass at $z=0.$ The blue curve shows the SFH of a typical main sequence galaxy for a halo of this mass, while the red curve shows a galaxy that experienced a quenching event at $z\approx1$ that shut down its specific star formation rate to negligible levels. The behavior of the quenching function, $F_{\rm q}(t),$ is shown in the middle panel of Figure~\ref{fig:sfh_ages}, which demonstrates the meaning of the $\qdrop$ parameter that controls the severity of the quenching event at time $t_{\rm q};$ the quenching function is implemented in terms of a triweight error function (see Appendix~\ref{sec:tdubs}). The bottom panel of Fig.~\ref{fig:sfh_ages} compares the distribution of stellar ages of these same two galaxies, with the solid and dashed curves showing results for different times of observation as indicated in the legend. Even though the top panel of Fig.~\ref{fig:sfh_ages} shows that the SFR of the main-sequence galaxy increases monotonically, the {\it specific} star formation rate is a decreasing function of time, and so by comparing dashed to solid curves of the same color in the bottom panel, we see that the composite stellar population ages as the galaxy evolves at late times.

\begin{figure}
\includegraphics[width=8cm]{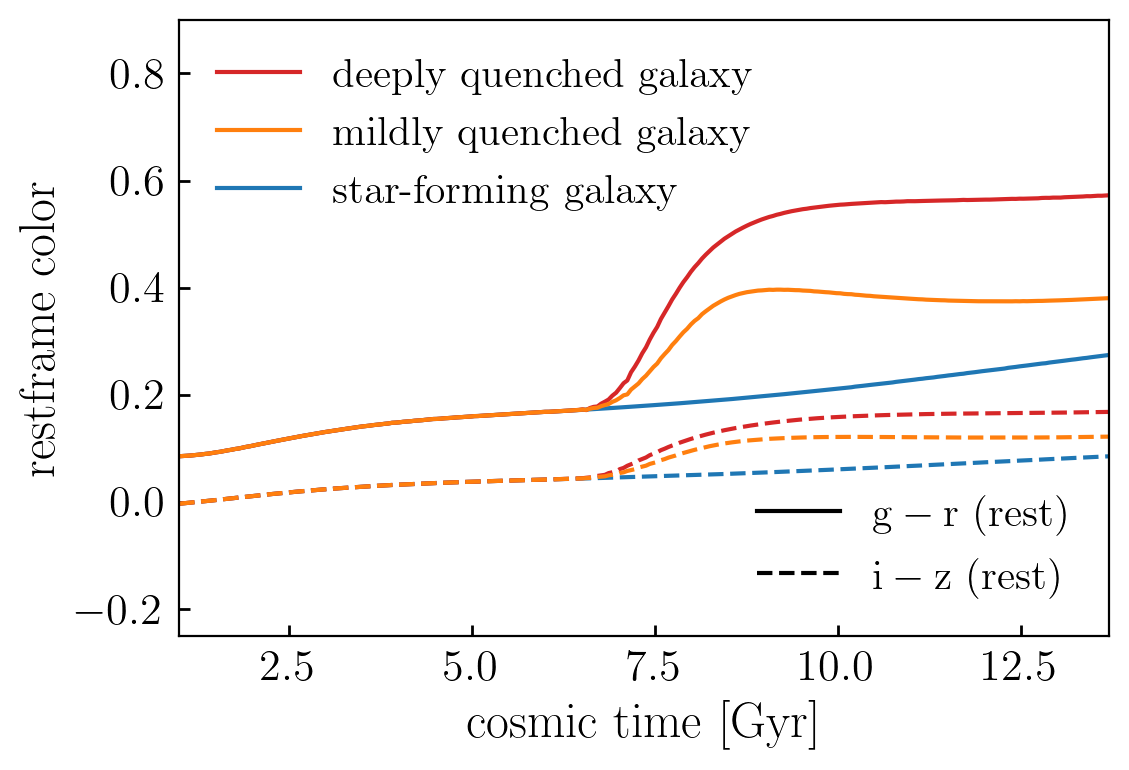}
\includegraphics[width=8cm]{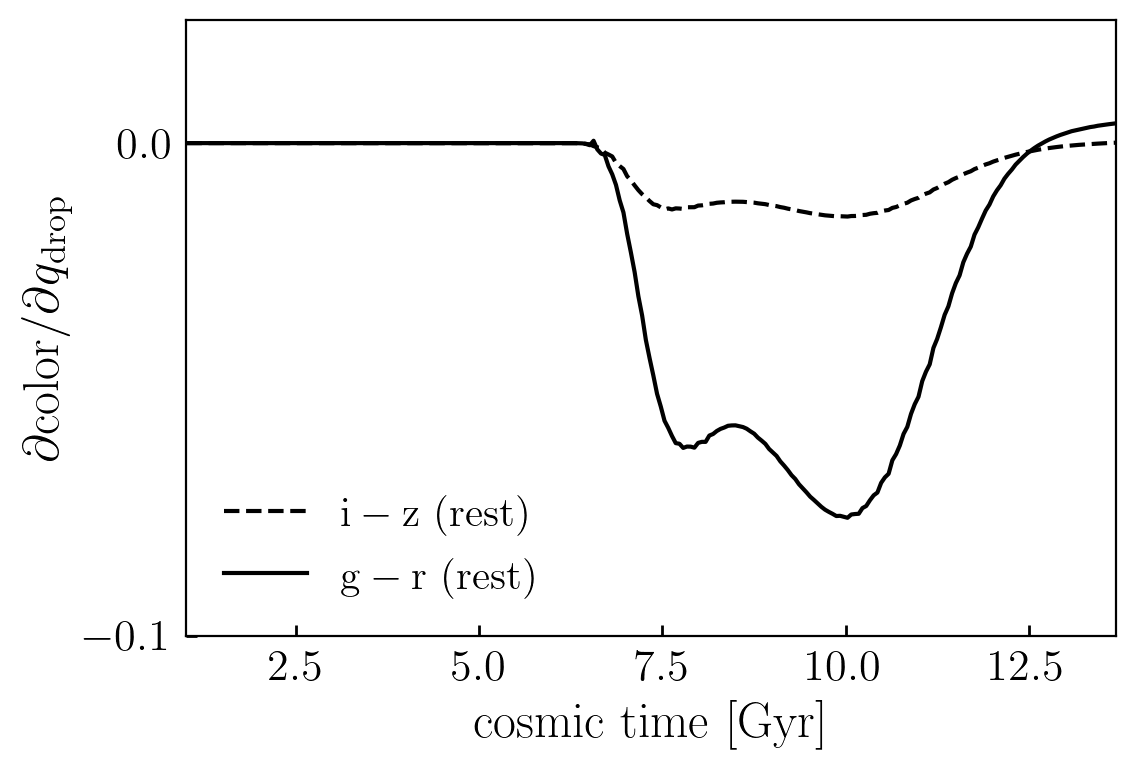}
\caption{{\bf SFH-dependent colors and gradients}. The top panel shows the history of the rest-frame colors of these same three galaxies appearing in Figure~\ref{fig:sfh_ages}, with colors observed through different filters as indicated in the legend. The bottom panel shows the gradient of the color of the quenched galaxy with respect to the parameter $\qdrop$ that describes the severity of the quenching event. Larger values of $\qdrop$ correspond to less extreme quenching events, and so the y-axis in the bottom panel is negative in the neighborhood of the quenching time, $t_{\rm q},$ since increasing $\qdrop$ produces a galaxy that is less quenched, and is thus generally bluer. We note that the gradient in the bottom panel was not calculated based on finite differencing methods, but was instead computed with the JAX autodiff library. See text for details.}
\label{fig:sfh_derivs}
\end{figure}

In the top panel of Figure~\ref{fig:sfh_derivs}, we show the history of the broad-band colors of these same two galaxies, plotted as a function of the time at which the galaxy is observed. We calculated g-r and r-i colors in the rest-frame according to Eq.~\ref{eq:restmag}, using transmission curves that mimic the filters of the Rubin Observatory Legacy Survey of Space and Time \citep[LSST, ][]{lsst_science_book}. We can see in Fig.~\ref{fig:sfh_derivs} that the colors of the galaxy redden with time; this is consistent with the aging of the stellar population of the galaxy visible in the bottom panel of Fig.~\ref{fig:sfh_ages}. The bottom panel of Fig.~\ref{fig:sfh_derivs} shows the gradient of the color history of the quenched galaxy only with respect to the parameter $\qdrop,$ defined as $$\qdrop\equiv\log_{10}\dot{M}_{\star}(t_{\rm q}) -\log_{10}\dot{M}_{\star}^{\rm ms}(t_{\rm q}),$$ where $t_{\rm q}$ is the first moment at which the quenching function $F_{\rm q}(t)$ reaches its lowest point (see the middle panel of Fig.~\ref{fig:sfh_ages} for a visual demonstration).\footnote{We note that in \dstar, $F_{\rm q}(t)$ can be non-monotonic because the parameterization of the model captures the possibility that some quenched galaxies experience rejuvenated star formation, but here we only demonstrate an example of permanent quenching.} 

When computing the gradient shown in the bottom panel of Fig.~\ref{fig:sfh_derivs}, for the fiducial point in parameter space, $\psfh,$ we use the star formation history of the quenched galaxy shown with the red curves in Fig.~\ref{fig:sfh_ages}, for which $\qdrop=-2.5;$ larger values of $\qdrop$ correspond to less extreme quenching events, and so the gradient of the galaxy color is negative in the neighborhood of $t_{\rm q},$ since increasing $\qdrop$ corresponds to a less quenched, and thus generally bluer galaxy. 

\subsection{Metallicity}
\label{subsec:metals}
In the previous section, we described the techniques required to make predictions for galaxy photometry that are differentiable with respect to star formation history parameters, and illustrated these techniques in Figure~\ref{fig:sfh_derivs} with a particular gradient of the rest-frame color history of a galaxy. When calculating galaxy photometry in \S\ref{subsec:ages}, we made the simplifying assumption that all the stars in the galaxy have the same metallicity; from this assumption, Equation~\ref{eq:cspsum} follows from the discretization of Equation~\ref{eq:csp}. We will now relax this assumption, and instead suppose that the metallicity of the stars in a galaxy are described by a parameterized Metallicity Distribution Function (MDF), $\probmet(Z).$

For a composite stellar population with a diversity in metallicity, the discretization of Equation~\ref{eq:cspsum} becomes
\beq
\label{eq:full_mdf}
\lcsp(\lambda\vert t) &= &\sum_{i, j}\lssp(\lambda\vert\tage^{i}, Z^{j})\\  && \times  P_{\rm SSP}(\tage^{i}, Z^{j}\vert t)\nonumber.
\eeq
In Eq.~\ref{eq:full_mdf}, the quantity $P_{\rm SSP}$ encodes the joint dependence of the SSP weights upon both the star formation history parameters, $\psfh,$ and the parameters regulating chemical evolution, $\pmet.$ The generalization of Equation~\ref{eq:cspsum} to Equation~\ref{eq:full_mdf} folds in an extra axis of summation over a new dimension in the data block storing the SEDs of the SSPs; by itself, this poses no technical obstacle for maintaining our autodiff-based implementation with JAX, and so the ability to calculate gradients of $\lcsp$ with respect to $\pmet$ rests on the differentiability of the MDF, $\partial P_{\rm SSP}/\partial\pmet.$

In principle, differentiable MDF implementations are possible even for codes such as \href{https://github.com/bretthandrews/flexCE}{{\tt flexCE}} \citep{andrews_weinberg_etal17_flexce} or \href{https://github.com/giganano/VICE}{{\tt VICE}} \citep{johnson_weinberg_2020_vice} that numerically solve systems of differential equations in which the influence of $\psfh$ and $\pmet$ are coupled, since JAX now supports several algorithms for numerically integrating ODEs. We discuss this future extension of DSPS in \S\ref{subsec:limits}, but for our present purposes, we will  demonstrate the differentiable influence of metallicity on SEDs using two simpler MDFs. In \S\ref{subsubsec:simple_metals}, we illustrate gradients of metallicity parameters for an MDF that is a clipped Gaussian with uncorrelated scatter, and in \S\ref{subsubsec:agedep_metals} we show gradients of an MDF that empirically captures correlations between stellar metallicity and age.

\subsubsection{Uncorrelated Metallicity Distribution Function}
\label{subsubsec:simple_metals}
The single-metallicity assumption used in \S\ref{subsec:ages} essentially assumes the MDF is a delta function centered on one of the particular metallicities for which $\lssp(\lambda)$ has been pre-computed. As a simple generalization, in this section we consider an MDF characterized by a clipped Gaussian distribution with scatter that is uncorrelated with any of the other parameters in the SPS model. Under this assumption, the two-dimensional probability distribution $P_{\rm SSP}(\tage, Z\vert\psfh, \pmet)$ becomes separable into two one-dimensional PDFs, $\probsfh(\tage\vert\psfh)$ and $\probmet(Z\vert\pmet).$ In the equation below, we will rewrite Eq.~\ref{eq:cspsum} for this MDF, now explicitly including the dependence upon the SPS model parameters for clarity:
\beq
\label{eq:simple_mdf}
\lcsp(\lambda\vert  t, \psfh, \pmet) &=& \Mstar(t)\cdot\sum_{i, j}\lssp(\lambda\vert\tage^{i}, Z^{j})  \\
&&\times\probsfh(\tage^{i}\vert t,\psfh)\cdot\probmet(Z^{j}\vert\pmet)\nonumber.
\eeq
From Eq.~\ref{eq:simple_mdf}, we can see that the ability to calculate exact gradients $\partial\lcsp/\partial\pmet$ rests on the differentiability of the MDF, $\partial\probmet/\partial\pmet.$

\begin{figure}
\includegraphics[width=8cm]{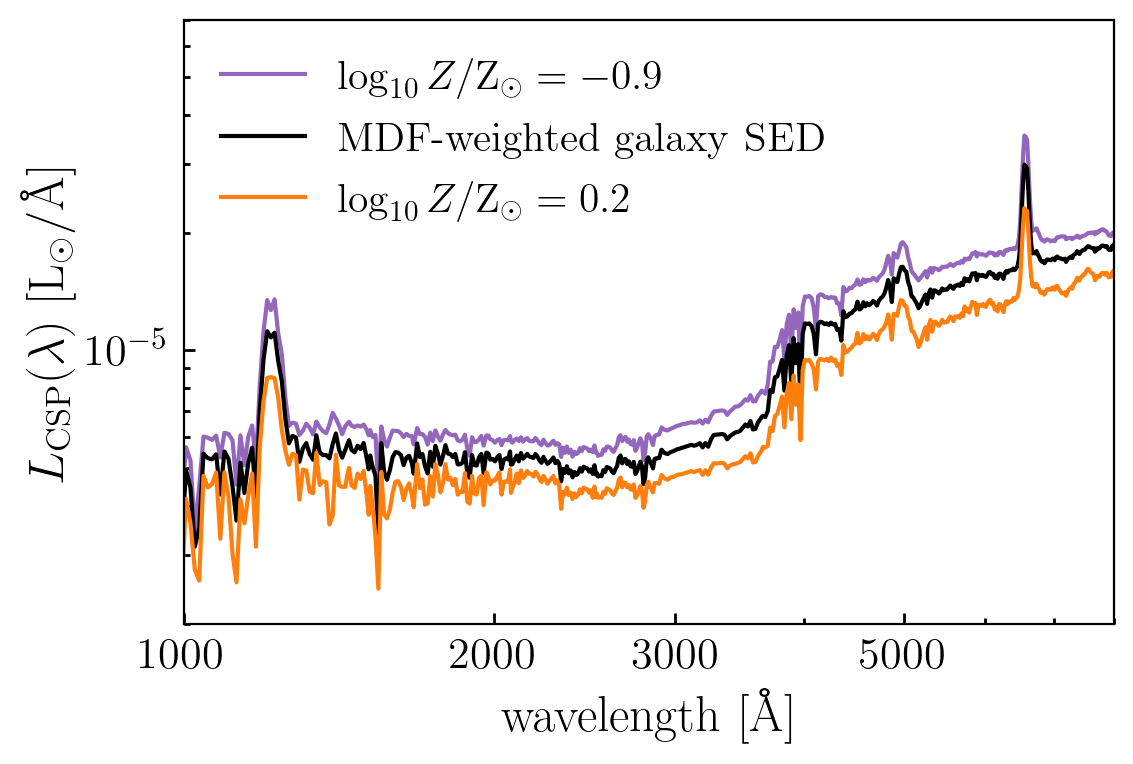}
\caption{{\bf Galaxy SED with simple MDF}. The black curve shows the SED of a main sequence galaxy at $z=0$ derived with a simple metallicity distribution function (MDF) centered at $\mumdf=\log_{10}Z/\Zsun=-0.3$ with uncorrelated scatter of $\sigmdf=0.25$ dex. The purple and orange curves show the SED of the SSPs with the smallest and largest metallicities that contribute to this MDF, respectively. The black curve is calculated as the MDF-weighted sum of all SSPs with metallicity in between the two bracketing cases.}
\label{fig:mdf_demo}
\end{figure}

We parameterize $\probmet$ using the triweight kernel, $\mathcal{T}(x\vert\mu,\sigma),$ which is very similar to a Gaussian distribution centered at $\mu$ with spread $\sigma$ that has been clipped to zero for $\vert{x-\mu}\vert\geq3\sigma$ (see Appendix~\ref{sec:tdubs} for details). In capturing the MDF with a triweight kernel, we will use $\log_{10}(Z/\Zsun)$ as our independent variable, so that $\probmet(Z\vert\pmet)$ is essentially a clipped log-normal distribution with uncorrelated scatter, and $\pmet=\left\{ \mumdf, \sigmdf\right\}.$ Figure~\ref{fig:mdf_demo} gives a simple demonstration of the SED of a galaxy derived from this triweight-based MDF.\footnote{Note that the SEDs we show in this section include a contribution from both starlight as well as emission from nebular gas, and so emission lines are visibly present in the spectra in Figure~\ref{fig:mdf_demo}. However, all SPS parameters controlling the physics of nebular emission are held fixed in this section; we postpone discussion of gradients with respect to the parameters of nebular emission until \S\ref{subsec:nebulae}.}  For the SFH of the galaxy shown in Fig.~\ref{fig:mdf_demo}, we used the same main sequence galaxy at $z=0$ shown in the figures in \S\ref{subsec:ages}; to calculate the SED of this example galaxy, we assumed an MDF centered at $\mumdf=\log_{10}(Z/\Zsun)=-0.3,$ with spread $\sigmdf=0.25;$ the SSP of the smallest and largest metallicities that contribute to the MDF are color-coded as indicated in the legend of Figure~\ref{fig:mdf_demo}.

\begin{figure}
\includegraphics[width=8cm]{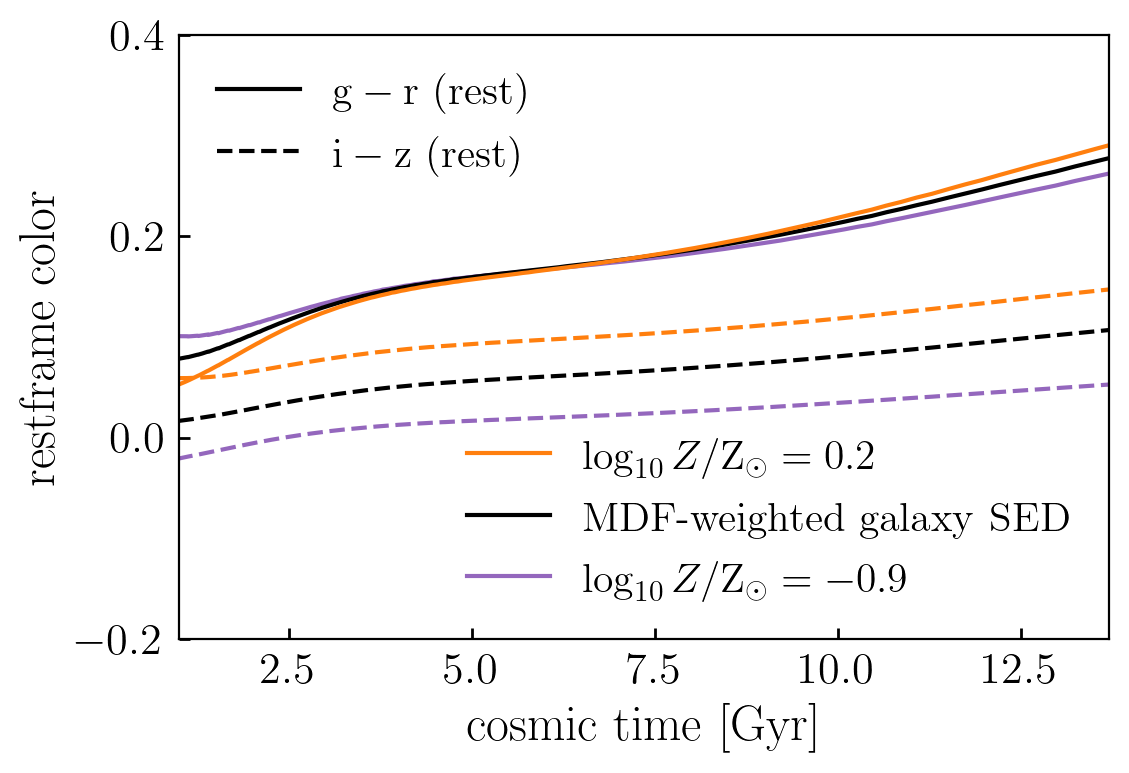}
\includegraphics[width=8cm]{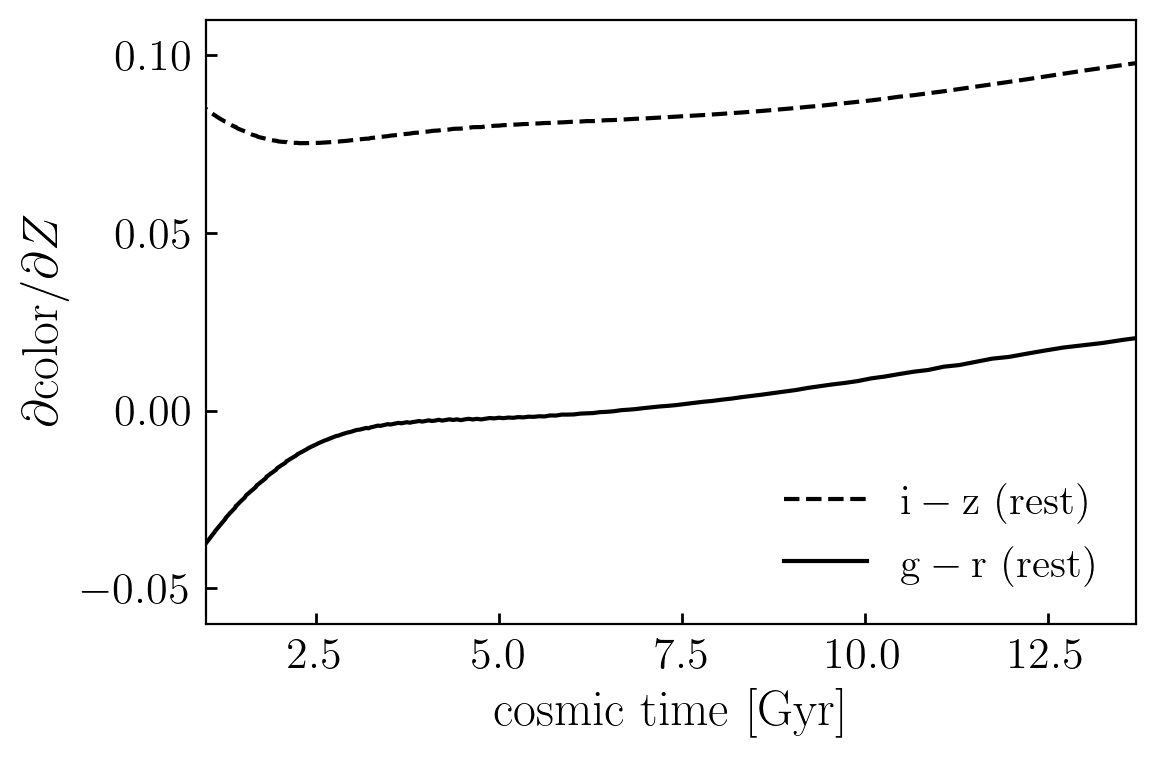}
\caption{{\bf Metallicity-dependent colors and gradients}. The top panel shows the history of the rest-frame colors of the same main sequence galaxy shown in Figure~\ref{fig:mdf_demo}. The bottom panel shows the gradient of the color with respect to $\mumdf,$ the mean metallicity of the log-normal MDF.}
\label{fig:met_derivs}
\end{figure}

In the top panel of Figure~\ref{fig:met_derivs}, we show the history of the broad-band colors of this same main sequence galaxy, plotted as a function of the time at which the galaxy is observed. The orange curve in Figure~\ref{fig:met_derivs} lies above the purple for most of cosmic time, so the higher-metallicity galaxy has redder rest-frame optical colors than the lower-metallicity galaxy, with the exception of the high-redshift behavior for the case of g-r color. This general trend is also apparent in the bottom panel of Figure~\ref{fig:met_derivs}, which shows the gradient of these color histories with respect to the parameter $\mumdf.$ 

\subsubsection{Age-Dependent Metallicity Distribution Function}
\label{subsubsec:agedep_metals}

\begin{figure}
\includegraphics[width=8cm]{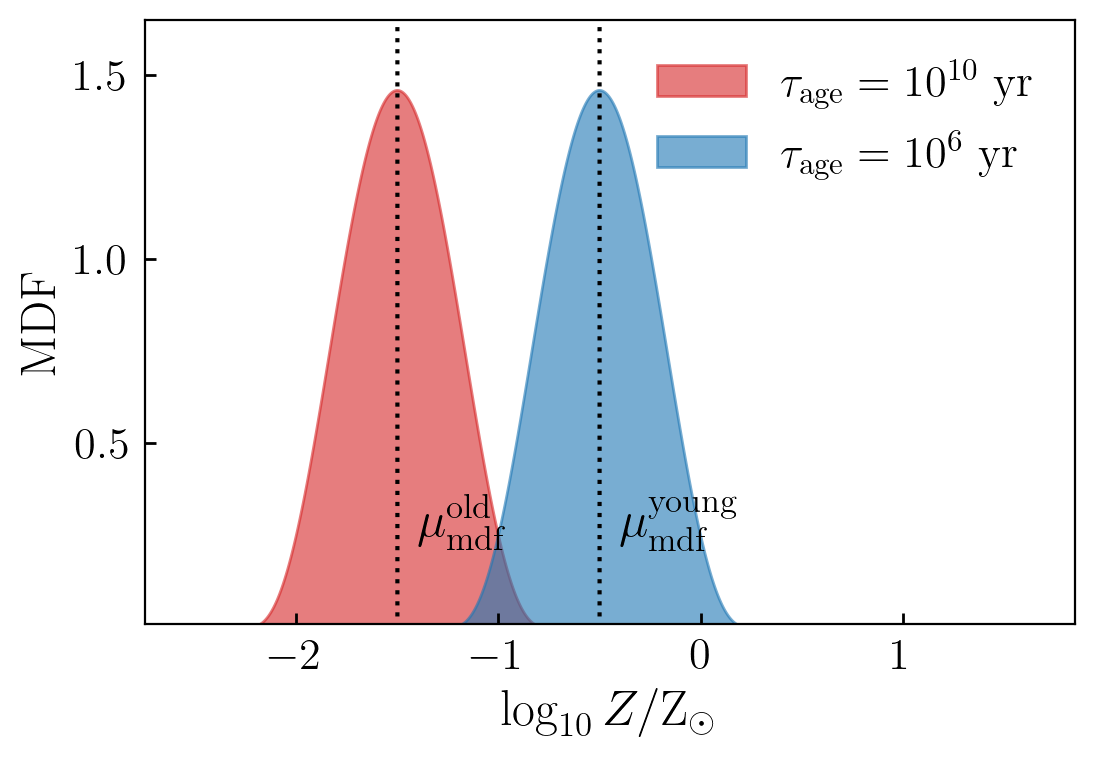}
\includegraphics[width=8cm]{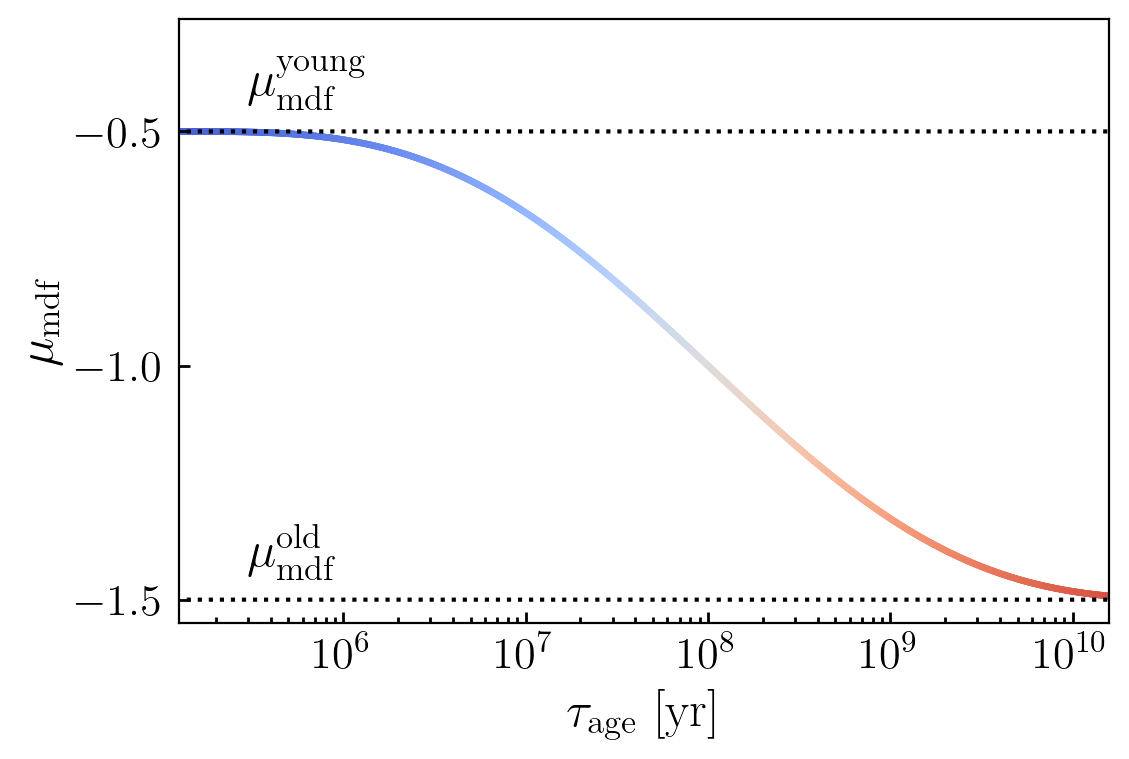}
\caption{{\bf Age-dependent MDF}. The two shaded histograms in the top panel show the metallicity distribution function (MDF) for stellar populations of different ages as indicated in the legend. At fixed $\tage,$ a triweight kernel is used to model the MDF as a clipped log-normal. The bottom panel shows the triweight cumulative kernel that smoothly transitions the central value of the MDF from a high-metallicity value for newly formed stars, $\mumdf^{\rm young},$ to a low-metallicity value for the oldest stars in the galaxy, $\mumdf^{\rm old}.$}
\label{fig:agedep_mdf_demo}
\end{figure}

The calculations in \S\ref{subsubsec:simple_metals} assumed that the MDF is a clipped log-normal distribution with uncorrelated scatter; this allowed us to simplify the expression for $\lcsp(\lambda),$ Eq.~\ref{eq:full_mdf}, by assuming that $P_{\rm SSP}(\tage,Z)$ is the product of two decoupled distributions, $\probsfh(\tage)$ and $\probmet(Z).$ In this section, we will relax this assumption with a simple empirical model for the MDF that captures possible correlations between stellar age and metallicity. We will maintain the use of a clipped log-normal distribution for metallicity, but allow for $\tage$-dependence of the center of the MDF: $$P_{\rm SSP}(\tage, Z)=\probsfh(\tage)\cdot\probmet(Z\vert\tage).$$
We will use a triweight cumulative kernel, $\terf(\log_{10}\tage),$ to capture the dependence of $\mumdf$ upon age, where $\terf(x)$ is the integral of a triweight kernel, as defined in Appendix~\ref{sec:tdubs}. Thus $\terf$ smoothly transitions the value of $\mumdf$ from an young-age value, $\mumdf^{\rm young},$ to an old-age value, $\mumdf^{\rm old},$ and at fixed $\tage,$ the MDF has the same triweight kernel shape used in \S\ref{subsubsec:simple_metals}. Figure~\ref{fig:agedep_mdf_demo} gives a visual demonstration of $\probmet(Z\vert\tage).$

\begin{figure}
\includegraphics[width=8cm]{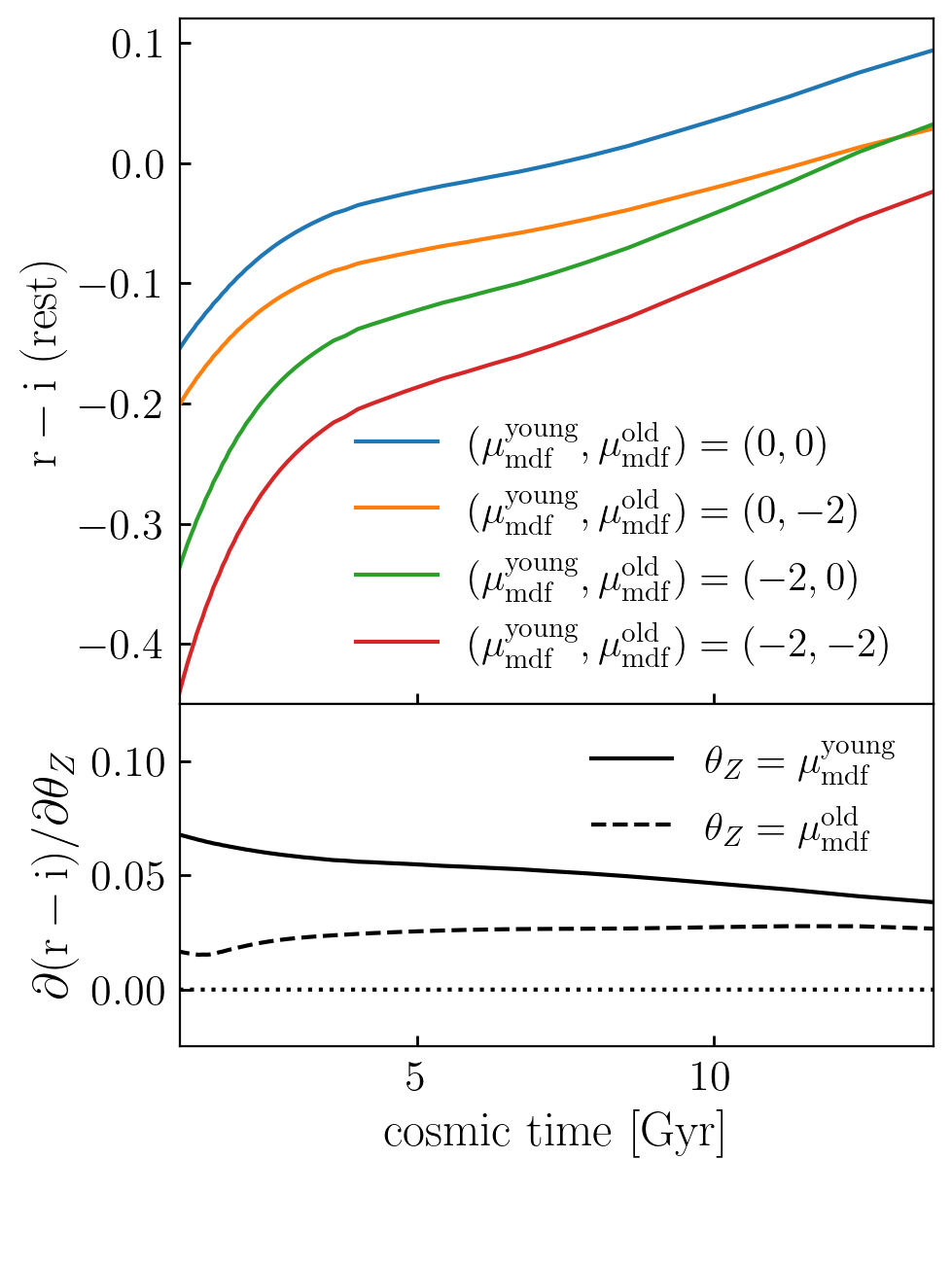}
\caption{{\bf Age-dependent MDF colors and gradients}. The top panel shows how the history of r-i color for stellar populations with different $\tage$-dependent metallicity as indicated in the legend. The bottom panel shows the autodiff-based gradient of r-i color with respect to each of the two parameters controlling the shape of MDF.}
\label{fig:agedep_mdf_grad}
\end{figure}

For this MDF, Equation~\ref{eq:simple_mdf} becomes:
\beq
\label{eq:agedep_mdf}
\lcsp(\lambda\vert t) &=& \sum_{i, j}\lssp(\lambda\vert\tage^{i}, Z^{j})\\
&\times&\probmet(Z^{j}\vert\tage^{i})\cdot\probsfh(\tage^{i}\vert t)\nonumber.
\eeq
In the top panel of Figure~\ref{fig:agedep_mdf_grad}, we show how the history of the broad-band galaxy color depends upon $\tage$-dependent metallicity. For a galaxy with the same main-sequence SFH studied above, each colored curve in the top panel of Fig.~\ref{fig:agedep_mdf_grad} shows the history of r-i color in the rest frame for a different combination of $\mumdf^{\rm young}$ and $\mumdf^{\rm old}.$ The bottom panel of Fig.~\ref{fig:agedep_mdf_grad} shows the gradient of r-i color with respect to each parameter. 

\subsection{Nebular Emission}
\label{subsec:nebulae}
For the results calculated in the previous sections, the spectra of the simple stellar populations, $\lssp(\lambda),$ include a contribution from both starlight as well as emission from nebular gas, but all parameters pertaining directly to the nebulae were held fixed, so that our SSPs could be described by a two-dimensional grid of $\left\{\tage^i, Z^j\right\}.$ In this section, we examine the differentiable influence on the composite SED of nebular emission parameters, $\pneb,$ focusing here on the impact of the ionization state of the nebular gas, parameterized by $\us=\log_{10}\Us.$ Adapting the differentiable techniques used in \S\ref{subsec:metals} to describe the MDF, we capture the effect of a probability distribution of $\probneb(\pneb)$ using a triweight kernel for the distribution of $\us:$
\beq
\label{eq:nebsed}
\lcsp(\lambda\vert  t, \psfh, \pmet, \pneb) &=& \Mstar(t)\cdot\sum_{i, j, k}\lssp(\lambda\vert\tage^{i}, Z^{j}, u^k) \nonumber \\
&\times&\probsfh(\tage^{i}\vert t,\psfh)\\
&\times&\probmet(Z^{j}\vert\pmet)\cdot\probneb(u^{k}\vert\pneb)\nonumber.
\eeq

\begin{figure}
\includegraphics[width=8cm]{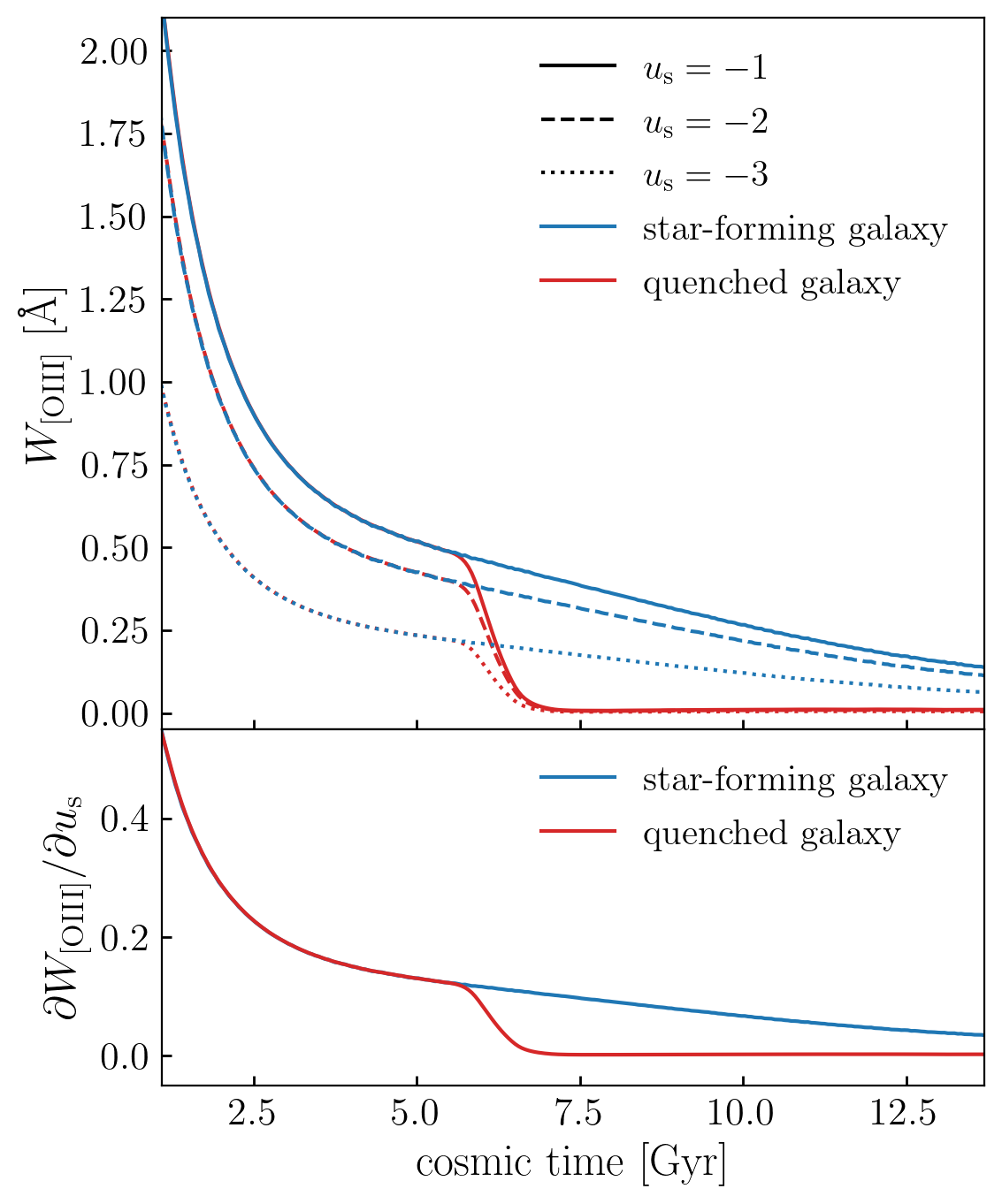}
\caption{{\bf Effect of nebular ionization on [OIII] emission}. We show the history of the equivalent width of the [OIII] emission line for same main-sequence and quenched galaxies shown in Fig.~\ref{fig:sfh_ages}. The top panel shows emission line histories for the three different values of $\us,$ as indicated in the legend; the bottom panel shows the gradient of the emission line history with respect to $\us.$}
\label{fig:OII_logU_gradient}
\end{figure}

We use {\tt python-fsps} to tabulate our collection of SSPs, $\lssp(\lambda\vert\tage^{i}, Z^{j}, u^k),$ and we use Equation~\ref{eq:nebsed} together with the techniques described in \S\ref{subsec:photometry} to calculate how emission line strength varies as a function of $\us.$ Figure~\ref{fig:OII_logU_gradient} shows the results of our calculations. For the [OIII] emission line at $\lambda=5000\Angstrom,$ we plot the equivalent width, $W_{\rm [OIII]},$ as a function of cosmic time for the same star-forming and quenched galaxies defined in \S\ref{subsec:ages}. The top panel of Fig.~\ref{fig:OII_logU_gradient} shows emission line histories for the three different values of $\us$ indicated in the legend. As discussed in \S\ref{subsec:ages},  the specific star formation rate of this galaxy is a decreasing function of time, and so the strength of [OIII] emission decreases as the galaxy evolves. We also see that the time $\qtime$ leaves a strong signature on the emission line history of the quenched galaxy; this is sensible, since [OIII] emission comes from the HII regions surrounding young stars, and the SFR of the quenched galaxy is shut down by the time $\qtime$ (see Figure~\ref{fig:sfh_ages}). 

The bottom panel shows the gradient of the emission line history, using $\us=-2$ as our fiducial point in parameter space, with scatter in $\sigma_{\us}=0.25.$ The positive sign of the gradient is sensible, since more highly ionized gas produces stronger [OIII] emission; for the red curve, the dropping of the gradient to zero after $\qtime$ is also expected, since the ionization parameter $\us$ has no appreciable effect on the composite SED after nearly all of star formation has been shut down.

\subsection{Dust Attenuation}
\label{subsec:attenuation}
Some of the starlight and nebular emission of a galaxy is obscured by dust, and the associated reduction in flux is captured by the attenuation curve, $\Alam,$ defined as
\beq
\label{eq:attenuation}
L_{\rm obs}(\lambda) \equiv \Fatt(\lambda)\times\lcsp(\lambda),
\eeq
where $\Fatt(\lambda)=10^{-0.4\Alam}.$
For the functional form of $\Alam$ presented here, we use the same parameterization as in \citet{salim_etal18}, defined as follows:
\begin{eqnarray}
\label{eq:salim}
\Alam = \frac{\Av}{4.05}\cdot\klam\\
\klam = k_0(\lambda)\cdot\left(\frac{\lambda}{\lambda_V}\right)^\delta + D_{\lambda}\nonumber
\end{eqnarray}
where $\lambda_V=5500\Angstrom;$ the quantity $k_0$ is piecewise-defined to be:
\beq
\label{eq:kzero}
k_0(\lambda)\equiv
\begin{cases}
\klamcal, & \lambda < 0.15\micron\\
\klamleh, & \lambda > 0.15\micron
\end{cases}
\eeq
with $\klamcal$ taken from \citet{calzetti00}, and $\klamleh$ taken from \citet{leitherer_etal02}; finally, the quantity $D_\lambda$ is the UV attenuation bump, which we model based on a Drude profile defined as:
\beq
\label{eq:uvbump}
D_\lambda \equiv \frac{E_{\rm b}(\lambda\Delta\lambda)^2}{\left(\lambda^2 - \lambda_{\rm b}^2\right)^2 + (\lambda\Delta\lambda)^2},
\eeq
where $\Delta\lambda=350\Angstrom$ and $\lambda_{\rm b} = 2175\Angstrom$ are constants, and we take $E_{\rm b}=-1.9\delta + 0.85$ as in \citet{kriek_conroy13}. See the  \href{https://github.com/ArgonneCPAC/dsps/blob/main/dsps/attenuation_kernels.py}{ attenuation\_kernels} module in DSPS for implementation details.

\begin{figure}
\includegraphics[width=8cm]{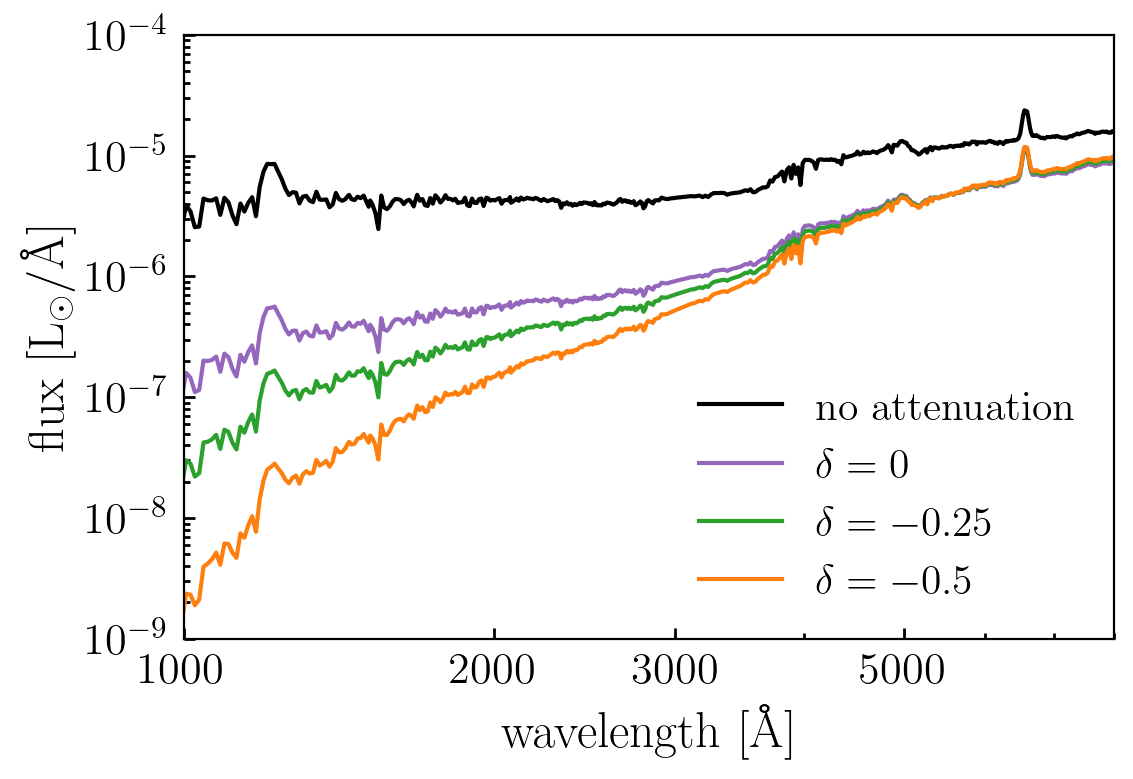}
\includegraphics[width=8cm]{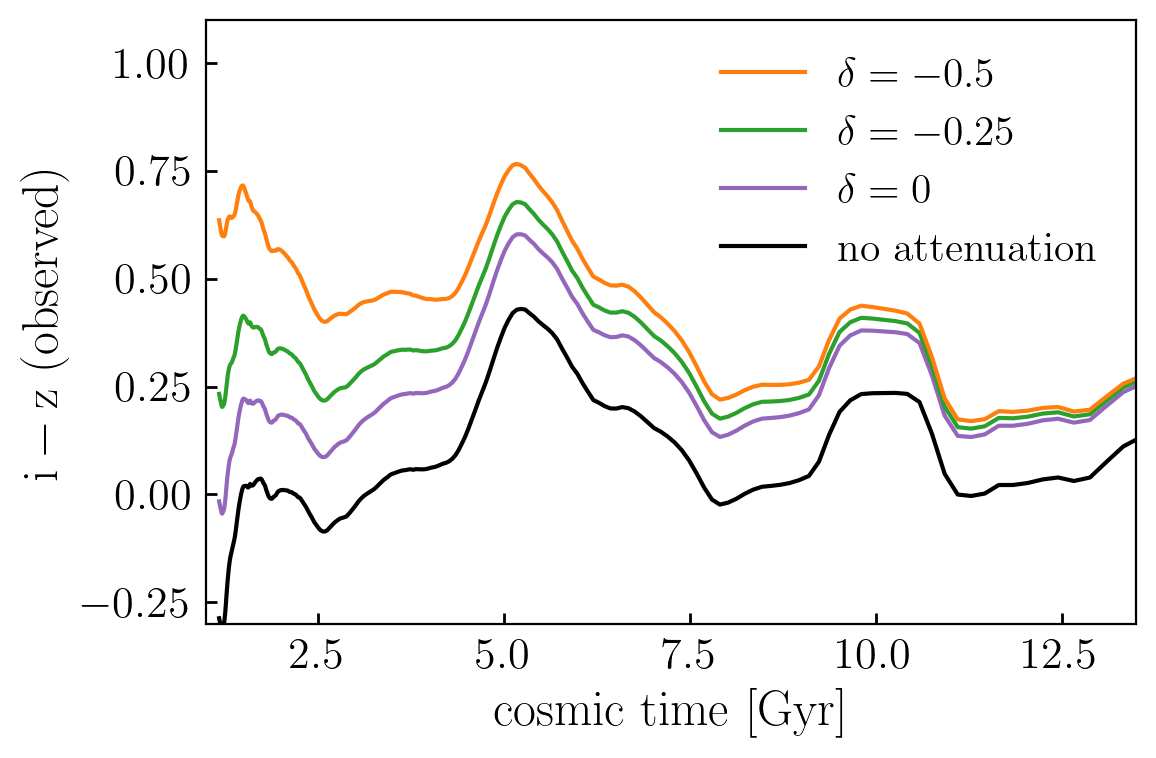}
\includegraphics[width=8cm]{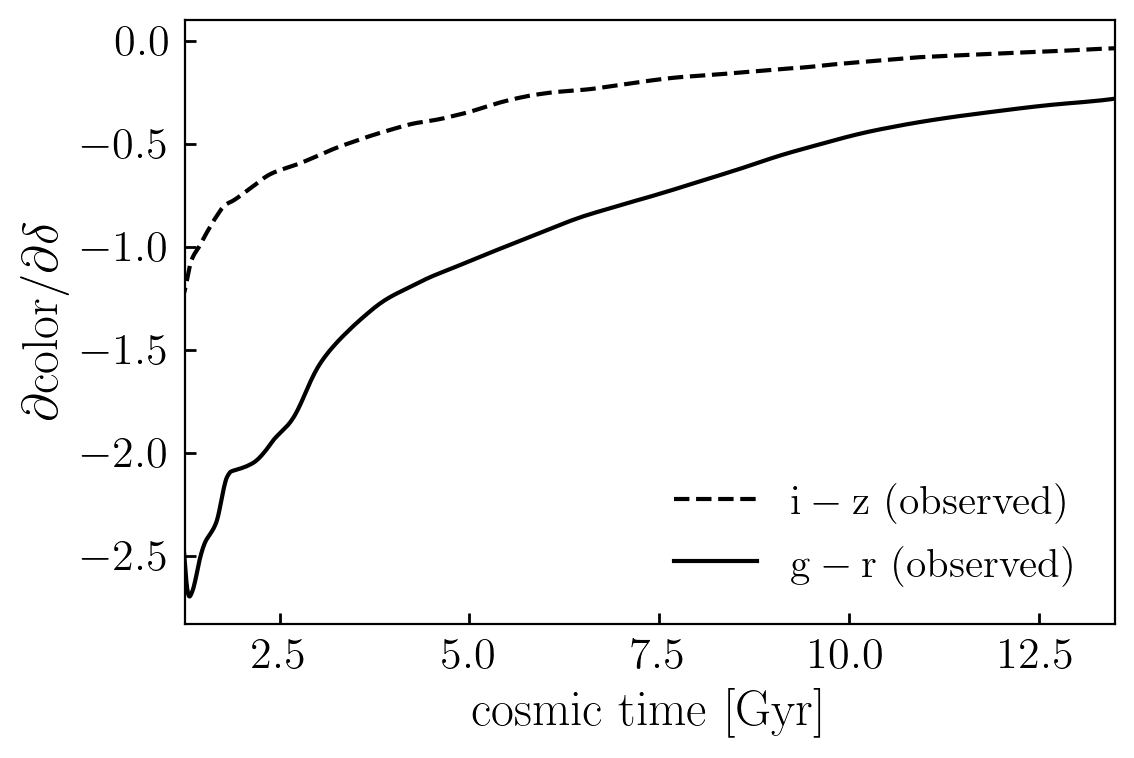}
\caption{{\bf Dust attenuation colors and gradients}. The top panel shows the SED of the same main sequence galaxy illustrated in Figure~\ref{fig:sfh_ages}; here we show this SED after applying attenuation curves, $\Alam,$ defined by different values of $\delta$ as indicated in the legend (see Eq.~\ref{eq:salim}). Each attenuation curve has $\Av$ equal to unity, except for the black curve, which shows an unattenuated SED with $\Av=0.$ The middle panel shows the i-z color history of the galaxy for different $\delta,$ and the bottom panel shows the autodiff-based gradient of color with respect to $\delta.$}
\label{fig:dust_derivs}
\end{figure}

We can see from Eqns.~\ref{eq:attenuation}-\ref{eq:uvbump} that the effect of $\Alam$ upon the observed SED is calculable analytically, and so in contrast to \S\ref{subsec:nebulae}, predicting the influence of dust attenuation on $\lcsp(\lambda)$ does not require expanding the dimension of the SSP grid. The top panel of Figure~\ref{fig:dust_derivs} shows $\lcsp(\lambda)$ for the same main sequence galaxy studied in the previous sections; different colored curves show SEDs with attenuation curves defined by different values of the parameter $\delta$ as indicated in the legend; each attenuation curve has $\Av$ equal to unity, except for the black curve, which shows an unattenuated SED with $\Av=0.$ From the top panel we can see that smaller (more negative) values of $\delta$ produce stronger levels of attenuation at shorter wavelengths, which follows from the shape of the functional form adopted for the attenuation curve.

In the middle panel of Fig.~\ref{fig:dust_derivs}, we show the i-z color in the observer frame for the same galaxy observed at different times. We can see that smaller values of $\delta$ correspond to redder colors, as expected from the top panel, and also that the color history of the galaxy has a complex time dependence produced by different spectral features that redshift in and out of the wavelength range of the i-z filter. In the bottom panel of Fig.~\ref{fig:dust_derivs}, we show the gradient of the color of the galaxy with respect to $\delta,$ which we have calculated for a fiducial value of $\delta_{\rm fid}=-0.25.$ The sign of the gradient is negative, again since larger values of $\delta$  produce optical colors that are less reddened by dust obscuration. 

\subsubsection{$\tage$-dependent attenuation}
\label{subsubsec:agedep_dust}

Some models of dust attenuation account for the natural physical expectation that the column depth and/or grain-size distribution of dust surrounding star-forming regions is distinct from the rest of the interstellar medium \citep[e.g.,][]{charlot_fall2000}. In this section, we adapt the attenuation model described above to incorporate this effect.

\begin{figure}
\includegraphics[width=8cm]{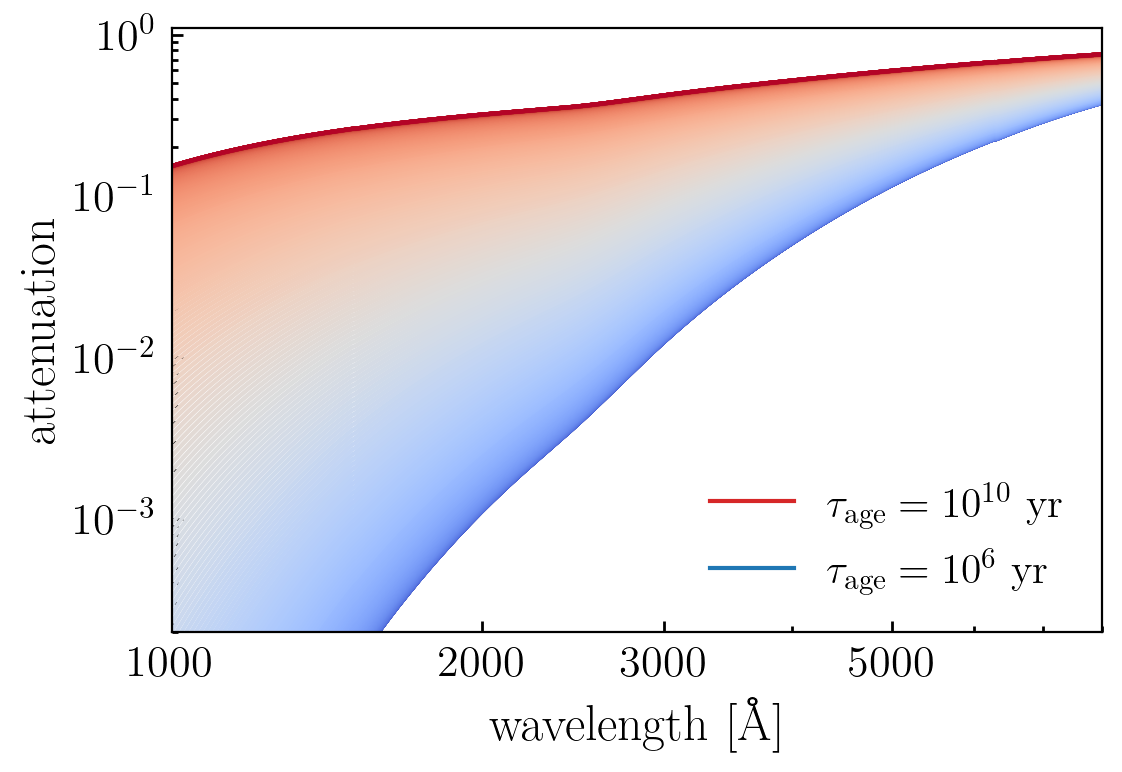}
\includegraphics[width=8cm]{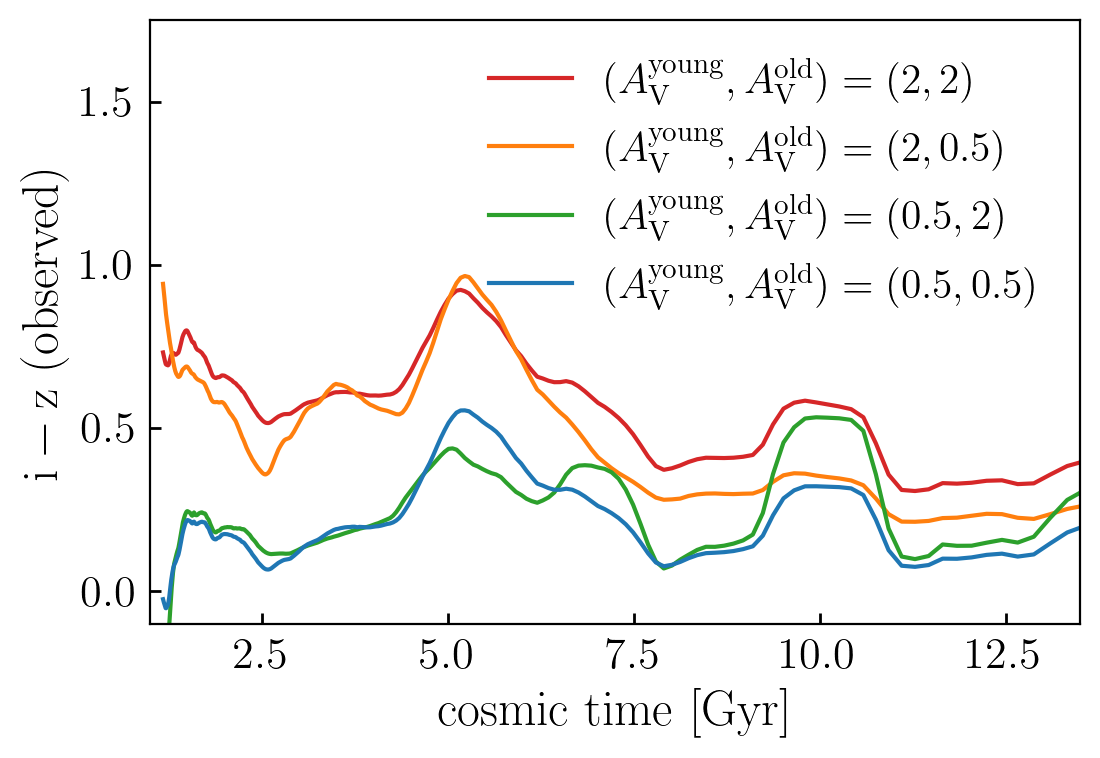}
\includegraphics[width=8cm]{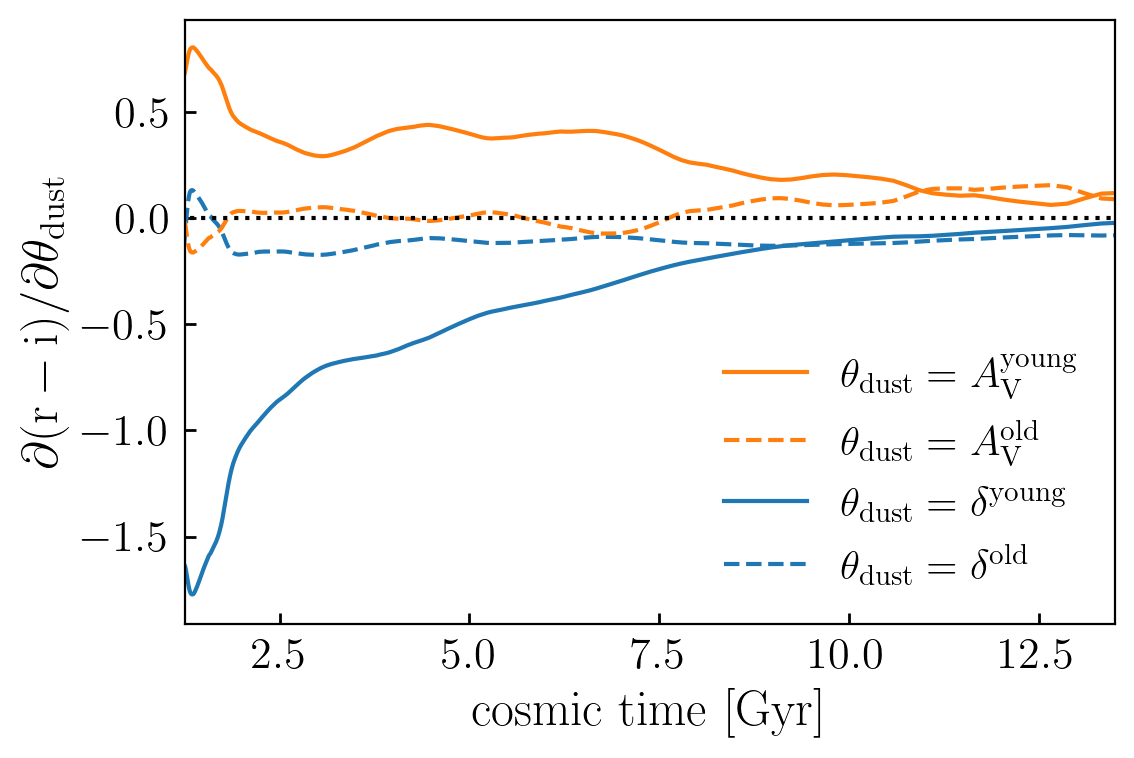}
\caption{{\bf Colors and gradients of $\tage$-dependent dust}. Analogous to Fig.~\ref{fig:dust_derivs}, only here $\Av(\tage)$ and $\delta(\tage)$ depend upon stellar age. 
The top panel shows the $\tage$-dependence of the attenuation curve, $\Fatt(\lambda\vert\tage),$ for the same main sequence galaxy illustrated in the previous sections. 
The middle panel shows the i-z color history of the galaxy for different combinations of $\Av^{\rm young}$ and $\Av^{\rm old}$, as indicated in the legend. The bottom panel shows the autodiff-based gradient of r-i color with respect to each of the four parameters controlling $\Av(\tage)$ and $\delta(\tage).$
}
\label{fig:agedep_dust_derivs}
\end{figure}

If the attenuation curve, $\Fatt(\lambda),$ depends upon stellar age, then our equation for the composite SED of a galaxy becomes:
\beq
\label{eq:agedep_dust_csp}
\lcsp(\lambda\vert  t, \psfh, \pmet) &=& \Mstar(t)\cdot\sum_{i, j}\lssp(\lambda\vert\tage^{i}, Z^{j})\nonumber\\
&\times&\Fatt(\lambda\vert\tage^i, \pdust) \\
&\times&\probsfh(\tage^{i}\vert t,\psfh)\cdot\probmet(Z^{j}\vert\pmet)\nonumber.
\eeq
For our model of $\Fatt(\lambda\vert\tage,\pdust),$ we use the same parameterization defined by Eqs.~\ref{eq:salim}-\ref{eq:kzero}, only we allow the parameters $\Av$ and $\delta$ to take on different values for young and old stars. For the functional form capturing the $\tage$-dependence, we use a triweight error function, $\terf(\log_{10}\tage),$ to smoothly transition the value of $\Av(\tage)$ and $\delta(\tage)$ from distinct asymptotic values for young and old stars (see Appendix~\ref{sec:tdubs} for the definition of $\terf$).\footnote{Note that in \S\ref{subsubsec:agedep_metals} we used essentially the same technique based on the triweight error function to capture the $\tage$-dependence of the metallicity distribution function.} 

Figure~\ref{fig:agedep_dust_derivs} illustrates how $\tage$-dependent attenuation impacts the color history of our fiducial main-sequence galaxy. The top panel shows how the attenuation curve, $\Fatt(\lambda\vert\tage),$ has joint dependence upon wavelength and stellar age. The middle panel of Fig.~\ref{fig:agedep_dust_derivs} shows the history of i-z color in the observer frame for the same main-sequence galaxy shown in Fig.~\ref{fig:dust_derivs}; color histories of galaxies with different combinations of $\Av^{\rm young}$ and $\Av^{\rm old}$ are shown with different curves, as indicated in the legend. In the bottom panel of Fig.~\ref{fig:agedep_dust_derivs}, we show the gradients of r-i color with respect to each of the four  parameters that control the $\tage$-dependence of dust attenuation. From both the middle and bottom panels we can see that broad-band optical color is more sensitive to the values of $\Av$ and $\delta$ in the dust surrounding young populations relative to old; this is sensible, since younger stars emit bluer light that is more susceptible to obscuration by dust relative to the redder light emitted by older stars. As discussed further in \S\ref{sec:discussion}, in future work we will leverage the flexibility of $\tage$-dependent dust models to capture the statistical correlations between attenuation and physical galaxy properties such as stellar mass and SFR, which are widely reported in observations \citep[see][for a review]{salim_narayanan_2020_dust_review}, and predicted by numerous hydrodynamical simulations of cosmological volumes \citep[e.g.,][]{hahn_etal21_eda}.

\section{Discussion}
\label{sec:discussion}

\subsection{Computational Benefits of JAX Implementation}
\label{subsec:jax}

In this paper, we have developed DSPS: a JAX-based implementation of many of the standard calculations of stellar population synthesis. The individual model evaluations in DSPS are quite efficient, as our code inherits the performance benefits of JAX, which is one of the highest-performance libraries used in contemporary deep learning.
Using the automated vectorization features of JAX, we have carried out a simple benchmarking experiment motivated by the simulation-based forward-modeling applications that are our principal science target. We performed the following operations in each iteration of our benchmarking experiment, repeating the sequence ten times and taking the median of the runtimes as our timing estimate:
\ben
\item We used \dstar to generate star formation histories for $10^5$ galaxies at $z=0;$
\item we used DSPS to calculate photometric fluxes through six LSST-like optical bands;
\item we applied a simple color cut to select galaxies with red colors in the resulting population;
\item we evaluated a histogram on $r$-band magnitude of the color-cut sample.
\een
This set of computational steps mimics the basic operations used in a simulation-based forward model of the luminosity function of a color-selected galaxy sample. On the \href{https://www.lcrc.anl.gov/systems/resources/swing/}{Swing} machine at Argonne, the computation with DSPS took 3.1 CPU-minutes on a single Intel Xeon E5 processor; the same computation requires $\sim15$ CPU-minutes using FSPS \citep{conroy_gunn_white_2009_fsps1,conroy_gunn_2010_fsps3}. Repeating the calculation using an NVIDIA A100 GPU on Swing required only 2.4 seconds of wall-clock time. Thus relative to standard SPS codes, for this calculation using DSPS provides a factor of $\sim5$ speedup on a CPU, and a factor of 375 on a modern GPU.\footnote{In this benchmarking exercise, we have used tabulations of $\lssp(\lambda)$ based on MILES \citep{falcon_barroso_etal11_miles_update}, but we note that these absolute runtimes depend on the wavelength resolution of the underlying collection of SSPs.} Even though FSPS is a mature library written in Fortran, these gains are significant even on a CPU due to the automatic vectorization features of JAX, which optimizes the memory layout of the computation in accord with the available resource. On a GPU these gains are particularly dramatic because we have formulated our SPS calculations in terms of linear algebra operations, and GPU hardware is extremely efficient at so-called SIMD calculations, in which the same instructions are applied to multiple data in parallel. We note that these vectorized operations can be memory-intensive, and so DSPS is better able to leverage these performance gains on high-performance computing machines with large memory resources.

There are numerous publicly available software libraries with GPU-efficient autodiff implementations besides JAX. For example, the authors in \citet{gully_santiago_blase_2022} have implemented a differentiable pipeline in PyTorch \citep{pytorch_citation_2019} for fitting the spectrum of an individual star; a variety of autodiff-based calculations in $\Lambda$CDM cosmography are included in the phytorch library \citep{Karchev_2022_phytorch}, which is also based on PyTorch; and the {\tt exoplanet} code \citep{dfm_exoplanet_joss_2021} for analyzing time-series astronomical data is based on pyMC3 \citep{pymc3}. The calculations presented in this paper could have alternatively been written in these or numerous other frameworks, as autodiff has become a common feature of contemporary deep learning libraries. Our choice to implement DSPS in JAX was primarily driven by a preference for the purely functional style of JAX, which we have found simplifies the task of integrating DSPS with other components of a larger scientific pipeline. This integration is facilitated by the fact that JAX itself is not a neural network framework, but rather is a library of composable transformations of n-dimensional arrays. The core transformations implemented in JAX are autodiff, vectorization, and parallelization, and DSPS uses these operations alone to reimplement its elementary SPS kernels. In ongoing work, we are using JAX to build neural network emulators of a few targeted bottlenecks of typical pipelines of SPS computations for galaxy populations. Integrating these networks into existing pipelines is essentially seamless, because neural networks in JAX are implemented as simply another composable transformation applied to the data.

The availability of gradient information is another benefit of the JAX-based implementation of DSPS. As discussed in \S\ref{sec:intro}, gradient-based optimization and inference algorithms substantially outperform other methods, particularly in high dimensions. Gradient availability also facilitates leveraging numerous recent developments in AI-enhanced inference \citep[e.g.,][]{zehgal_etal22_npe,wong_etal22_flowmc}, and advanced inference techniques such as these are becoming increasingly common in SED modeling \citep[e.g.,][]{hahn_desi_provabgs_2022,hahn_melchior_2022_npe,khullar_nord_etal2022_digs}. Gradient information is also extremely useful when conducting a sensitivity analysis; for example, tracing the trajectory of exact gradients simplifies the task of identifying degeneracy directions across the parameter space of a model, and the availability of second-order gradients enables methods using the Fisher information matrix.

We expect that the property of differentiability will play a key role in simulation-based forward models of the galaxy--halo connection. In such applications, the target data vectors are summary statistics of an entire galaxy population, whereas in this work we have only demonstrated gradients pertaining to individual galaxies. In forward-modeling applications targeting large-scale structure observables, we will use the techniques presented in \citet{hearin_etal21_shamnet} to propagate the gradients of individual SEDs through to the one-point estimators used to measure the luminosity function and the distribution of galaxy colors, as well as the two-point estimators of the clustering and lensing of galaxy samples selected by their photometry and/or emission lines. Coupling these techniques to the Diffstar model will enable cosmological simulation-based analyses of an entire population of galaxy SEDs and photometry, including predictions for the redshift-dependent spatial distribution of galaxy SEDs across linear and nonlinear regimes \citep[see][for further discussion]{alarcon_etal21}.

Direct reimplementation of SPS computations in JAX is not the only method by which differentiable SED predictions can be achieved. For example, in \citet{alsing_etal20_speculator}, the authors trained several neural networks (referred to collectively as SPECULATOR) to approximate the SED and photometry predictions of two different SPS models, achieving a surrogate function that is both differentiable and provides a factor of $10^3-10^4$ speedup relative to exact evaluation with standard SPS codes. In this approach to accelerating the computations of stellar population synthesis, an existing SPS library is treated as a black box that supplies training data quantifying how the galaxy SED responds to the input model parameters. The advantage of this approach is that the end result represents essentially the extreme limit of the computational efficiency that can be achieved for the specific model prediction that is emulated. The disadvantage is that the human effort required to identify the appropriate architecture and optimize the parameters of the network can be quite considerable, and the associated labor must be repeated each time there is even a small change to the configuration of the analysis. Since the calculations implemented in DSPS are not emulated, but are the same computations implemented in other SPS libraries, our performance gains derive from the one-time effort of our JAX-based reimplementation; for the same reason, SED calculations with DSPS use the exact SSPs and SFHs, and are not reliant on approximations based on principal components and Gaussian mixtures.

Of course, there is no conflict between the use of emulation methods and our reimplementation of SPS in an autodiff library. In fact, quite the opposite is true. In generating $\sim10^6$ points of training data for an SPS emulator, the performance gains quoted above translate into the reduction of wall-clock times from an entire day to only a few minutes on a GPU. Beyond this simple translation of our speedup factor into practical terms, the purely functional software design of the JAX library enables a shift to the workflow of emulation. Since the entirety of DSPS is implemented in JAX, it is straightforward to replace specific components of the prediction pipeline with a dedicated emulator. For example, in profiling DSPS as part of this work, we identified one of the dominant bottlenecks of our SED predictions to be the calculation of $P(\tage\vert\theta_{\rm sfh}),$ the probability distribution of stellar ages as a function of the SFH parameters (see \S\ref{subsec:ages} for details). This implies that a neural network emulator of $P(\tage\vert\theta_{\rm sfh})$ has potential to provide a considerable performance enhancement. We highlight that this alternative approach to emulation does not require sacrificing the flexibility of the analysis. Whereas an end-to-end photometry emulator must be retrained each time one varies the filters, by contrast, once a surrogate function for $P(\tage\vert\theta_{\rm sfh})$ has been trained, the performance gains apply directly to all downstream SED predictions. The focused nature of this approach to emulation also simplifies training, since simpler target functions are less challenging to emulate. 

\subsection{Current Limitations and Future Work}
\label{subsec:limits}

As described in \S\ref{sec:sps}, DSPS takes as its starting point a data block such as $\lssp(\lambda\vert\tage^{i},Z^{j})$ that represents the SEDs of a collection of simple stellar populations. All of the calculations demonstrated in \S\ref{sec:ingredients} were based on SSPs for which a fixed IMF had been assumed, so that each element of the input data block provides an interpolation table for the SED of a homogeneous population of stars with age, $\tage^{i},$ and metallicity, $Z^{j}.$ The differentiable techniques in DSPS can work equally well with SSP templates that include additional dependencies beyond metallicity and age; for example, the calculations in \S\ref{subsec:nebulae} included additional dependence of the SSPs on the ionization state of the nebular gas, $U_{\rm s},$ so that the data block of SSPs has an extra dimension, $\lssp(\lambda\vert\tage^{i},Z^{j},U_{\rm s}^{k}).$

Depending on the physics under consideration, it may or may not be necessary for the dependence of the SSP upon some parameter, $\theta,$ to appear in the input data block. Consider, for example, $\theta_{\rm dust},$ the parameters of the attenuation curves appearing in \S\ref{subsec:attenuation}; in this case, the effect on the SED is calculable analytically in JAX, and so there is no need for the data block of the input SSPs to include a discretized pre-tabulation of the effect of $\theta_{\rm dust}.$  For modeling ingredients in which it is straightforward to implement the effect on the SED directly in JAX, this is generally preferred, because with each new variable upon which the SSPs depend, there is a multiplicative increase of the memory footprint required by the SED calculation. In all cases, the required template spectra are not provided by DSPS, and so users of DSPS must acquire these from another library that supplies them. 

There is no strict need for differentiable SED calculations to rely on fixed-IMF SSPs; one could instead begin with an input data block that provides a discretized tabulation of single-star SEDs, $\lss(\lambda\vert,\tage^{i},Z^{j},\Mstar^{k}).$ 
Beginning with $\lss(\lambda)$ as the starting point creates the opportunity to forward model the effect of parameters encoding uncertainty in the IMF, or even stellar rotation and the fraction of stellar binaries. The additional complicating factor in this case is that while $\tage$ is the natural independent variable appearing in SPS calculations, for single-star spectra the natural variable is the equivalent evolutionary phase \citep[EEP, see, e.g.,][]{dotter16_mesa_isochrones}, and the $\tage$--EEP relationship is highly non-trivial. Publicly available libraries such as \href{https://github.com/timothydmorton/isochrones}{isochrones} \citep{morton_2015_isochrones} can be used to calculate how EEP depends upon $\tage$ for stars of different mass and metallicity; these calculations can in turn be used to train a targeted emulator for the $\tage$--EEP relationship, which is all that would be required to perform differentiable SED calculations using tabulations of $\lss(\lambda)$ as a starting point. While this is beyond the scope of the present paper, we will present ongoing results in this direction in a follow-up paper that extends the functionality of the DSPS library.

In \S\ref{subsec:metals}, we presented a simple model for stellar metallicity that captures both diversity in the chemical composition of the stars in a galaxy, as well as potential correlations between the metallicity and age of stars within a composite population. Detections of such correlations have been reported in observational data for both large extragalactic samples \citep{poggianti_etal01,gallazzi_etal05} as well as within the Milky Way \citep{hayden_etal15}, and are expected on basic physical grounds. Of course, due to the well-known age--metallicity degeneracy \citep{worthey_1994_age_metal_degeneracy}, it is challenging to disentangle these two influences on the SED, and so it may not be possible to constrain the chemical makeup of a galaxy beyond its average metallicity, particularly when the measurements are limited in resolution and/or wavelength range. Nonetheless, the model in \S\ref{subsec:metals} improves upon the widely-used assumption of constant metallicity, requires only one or two additional parameters, and comes with a practically negligible loss of computational efficiency.

Future extensions of DSPS will include physical models for the chemical evolution of a galaxy that complement the empirical approach taken in \S\ref{subsec:metals}. We are currently pursuing two distinct approaches to these extensions. The first is motivated by \citet{weinberg_etal17}, who showed that the differential equations governing the chemical evolution of a galaxy admit analytical solutions for a wide range of functional forms of star formation history, even when assumptions of instantaneous recycling are relaxed. The computational advantages of this analytical approach are attractive, particularly in light of the diversity in metallicity distribution functions (MDF) that can be captured by the solution space. In the second approach, we directly parameterize the MDF evolution of individual galaxies, seeking a functional form that is sufficiently flexible to capture what is seen in the merger trees of galaxies in hydrodynamical simulations. This essentially mirrors the effort in \citet{alarcon_etal21} to build the \dstar parameterization of star formation history, and has the advantage of relaxing the equilibrium assumptions used to derive the analytical solutions of the differential equations of chemical evolution.

Throughout this paper, we focused on the \dstar model for demonstrating autodiff-based gradients of galaxy SEDs; as shown in \citet{alarcon_etal21}, this model is flexible enough to give an unbiased description of the SFHs in the \um and \tng simulations. However, we note that \dstar is not as flexible as piecewise-defined models \citep[e.g.,][]{leja_etal19_howtosfh2}, since \dstar essentially imposes constraints on the shape of SFHs derived from basic scaling relations of the galaxy--halo connection. Since the influence of SFH on broad-band optical colors is relatively simple and largely insensitive to complex patterns of burstiness \citep{chaves_montero_hearin_2020_sbu1,chaves_montero_hearin_2021_sbu2}, then smooth models such as \dstar may be sufficient for purposes of predicting and interpreting the data from large cosmological imaging surveys. However, recent results indicate that the freedom offered by piecewise-defined models offers significant benefits over some widely-used parametric SFH forms \citep{lower_etal20}, and so it remains to be seen whether a smooth model such as \dstar can improve upon these limitations of traditional parametric forms. Fortunately, there is no technical obstacle to implementing a range of piecewise-defined models in a differentiable fashion; the behavior of such models is essentially defined by some form of interpolation on a grid, which can be naturally implemented with an autodiff library whether this operation is carried out with linear interpolation \citep[e.g.,][]{chauke_etal18,leja_etal19_3dhst,johnson_etal21_prospector}, or via an ML algorithm such as Gaussian Process \citep[as in][]{iyer_etal19}.

Some SPS models use truly non-parametric formulations of SFH that are defined by randomly sampling from a pre-computed library of SFHs supplied by a galaxy formation simulation \citep{finlator_etal07,Pacifici2012,pacifici_etal15}. This class of models can also be formulated in a differentiable fashion by adapting the GalSampler technique \citep{hearin_etal20_galsampler}: the simulated SFHs in the library are selected with a parametrized function that controls the PDF used in the weighted sampling. We are currently developing this SFH model based on galaxy libraries generated with the Galacticus SAM \citep{benson_galacticus_2012}, and will explore this further in future work.

The current version of DSPS does not include several physical ingredients that are commonly implemented in other widely-used libraries for stellar population synthesis. For example, the models presented in \S\ref{subsec:attenuation} only capture how dust absorbs optical and UV starlight, but not how the absorbed energy is reemitted by the dust, and so DSPS does not yet have capability to make physically realistic predictions for the SEDs of dusty star-forming galaxies in the infrared. Exploring these effects is beyond the scope of this paper, since our present focus is on demonstrating the scientific potential of autodiff-based implementations of stellar population synthesis, but we will include these and other physical ingredients in future releases of DSPS.

\section{Summary}
\label{sec:summary}
We conclude by summarizing our primary results:
\begin{enumerate}
    \item We have developed DSPS, \url{https://github.com/ArgonneCPAC/dsps}, a stellar population synthesis code written in the JAX library for automatic differentiation. Our software is available for installation with conda or pip.
    \item In applications of simulation-based forward modeling the galaxy--halo connection, DSPS improves upon the computational performance of standard SPS codes by a factor of 5 on a CPU, and by over a factor 300-400 on a GPU.
    \item We have detailed a set of techniques for formulating models of SPS in a manner that makes the predictions naturally differentiable with autodiff, demonstrating worked examples of gradients of the parameters of models of star formation history in \S\ref{subsec:ages}, metallicity in \S\ref{subsec:metals}, nebular emission in \S\ref{subsec:nebulae}, and dust attenuation in \S\ref{subsec:attenuation}.
    \item In Appendix~\ref{sec:remnants}, we have developed new fitting functions for the fraction of stellar mass that survives as a function of time, $\fsurv(t),$ and also the fraction of a population locked up in stellar remnants, $\frem(t);$ our fitting functions accurately capture the IMF-dependence of $\fsurv(t)$ and $\frem(t),$ and so their standalone implementations in DSPS may be useful in other SPS applications.
\end{enumerate}

\section*{Acknowledgements}

Special thanks to Ben Johnson, whose active support of FSPS has created a public record on GitHub that has been an invaluable resource in developing this work. APH thanks José Feliciano for Feliz Navidad.

We thank the developers of {\tt NumPy} \citep{numpy_ndarray}, {\tt SciPy} \citep{scipy}, Jupyter \citep{jupyter}, IPython \citep{ipython}, scikit-learn \citep{scikit_learn}, JAX \citep{jax2018github}, conda-forge \citep{conda_forge_community_2015_4774216}, and Matplotlib \citep{matplotlib} for their extremely useful free software. While writing this paper we made extensive use of the Astrophysics Data Service (ADS) and {\tt arXiv} preprint repository. 

Work done at Argonne was supported under the DOE contract DE-AC02-06CH11357. We gratefully acknowledge use of the Bebop cluster in the Laboratory Computing Resource Center at Argonne National Laboratory. This work was performed in part at Aspen Center for Physics, which is supported by National Science Foundation grant PHY-1607611. APH and AB acknowledge support from NASA under JPL Contract Task 70-711320, “Maximizing Science Exploitation of Simulated Cosmological Survey Data Across Surveys.”

\section*{Data Availability}
Data underlying this article is publicly available at the DSPS code repository on github, \url{https://github.com/ArgonneCPAC/dsps}.

\bibliographystyle{mnras}
\bibliography{bibliography}

\appendix
\renewcommand{\thefigure}{A\arabic{figure}}

\counterwithin{figure}{section}
\renewcommand{\thefigure}{A\arabic{figure}}
\section{Fitting Functions for Stellar Lifetimes and Remnants}
\label{sec:remnants}

\begin{figure}
\includegraphics[width=8cm]{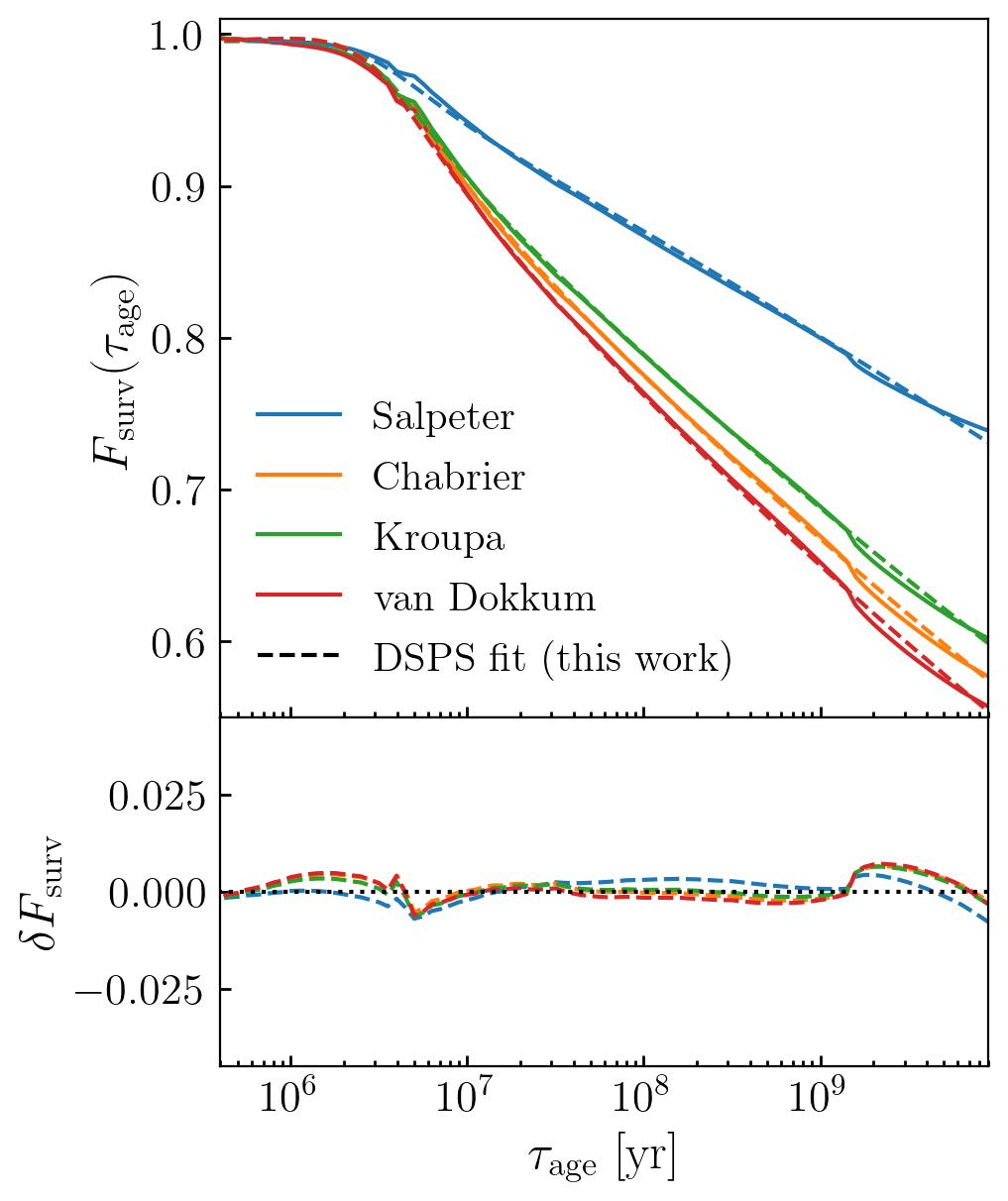}
\caption{{\bf Surviving stellar mass fraction}. For a simple stellar population that formed in a single burst, the quantity plotted on the y-axis, $\fsurv(\tage),$ describes the fraction of the initial mass of the population that survives until the time $\tage$ following the burst. Due to the $\Mstar$-dependence of stellar lifetimes, $\fsurv(\tage)$ depends on the IMF in a manner shown by the different colored curves in the figure. Solid curves show results calculated with the FSPS library; dashed curves show the corresponding fitting function used in DSPS. The bottom panel shows the residual error in the approximation supplied by the fitting function.}
\label{fig:msurv_imf}
\end{figure}

\begin{figure}
\includegraphics[width=8cm]{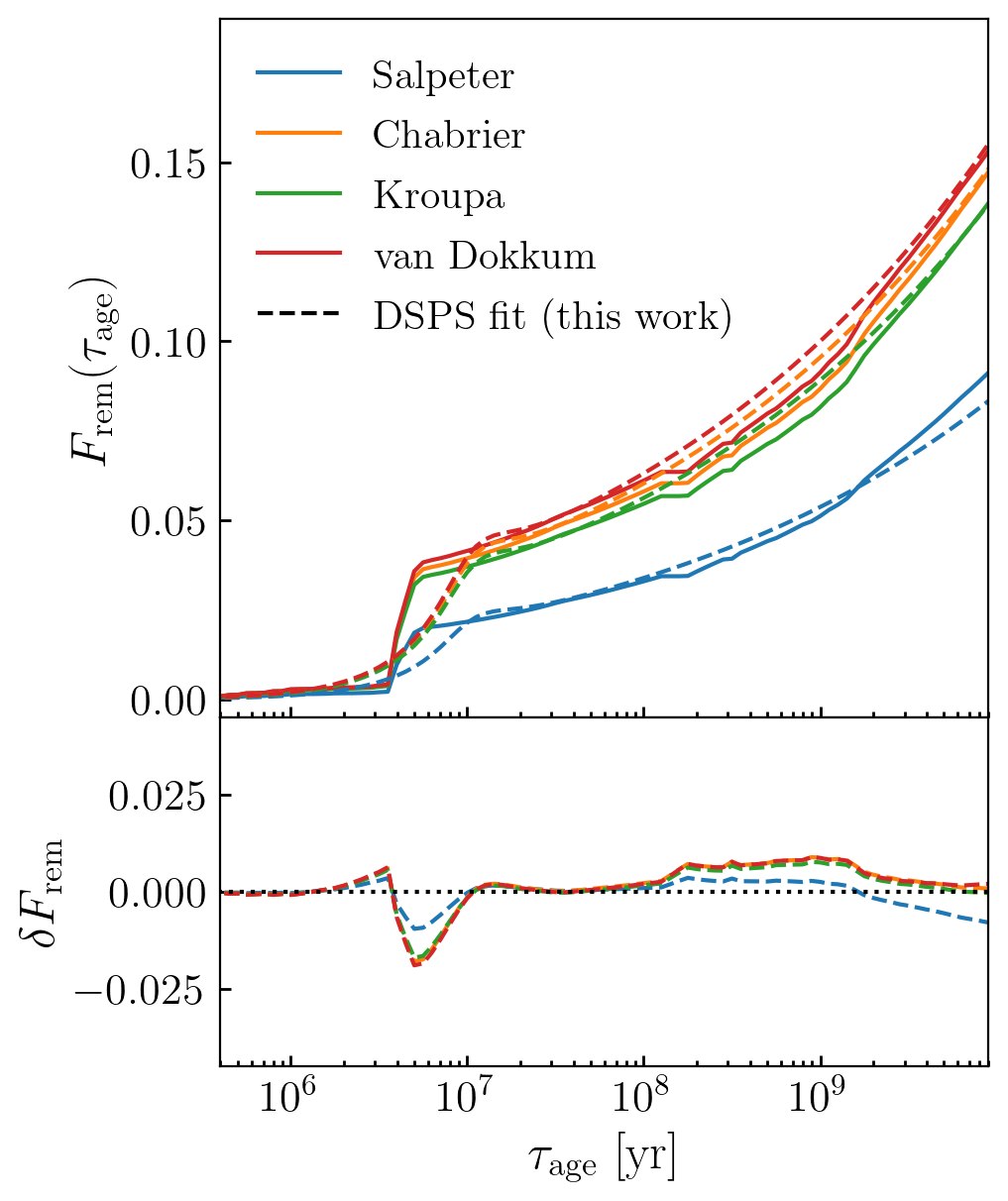}
\caption{{\bf Remnant mass fraction}. The quantity $\frem(\tage)$ describes the fraction of the initial mass of a simple stellar population that has been transformed into stellar remnants at the time $\tage.$ As in Figure~\ref{fig:msurv_imf}, different colored curves show results for different assumed IMFs as indicated in the legend; solid curves show results derived with FSPS; dashed curves show the corresponding fitting functions implemented in DSPS; residual errors are shown in the bottom panel.}
\label{fig:mrem_imf}
\end{figure}
The lifetime of a star depends sensitively on its initial mass. A massive O star with $\Mstar\gtrsim50\Msun$ will explode as a core-collapse supernova after a few million years, whereas an M star with mass $\Mstar\approx0.1\Msun$ will continue to burn hydrogen for trillions of years, far longer than a Hubble time. Thus after a time $\tage$ following some initial burst of star formation, only some fraction of the stars that form continue to survive, $\fsurv(\tage),$ and the remaining mass is returned to the interstellar medium. Due to the mass-dependence of stellar lifetimes, $\fsurv(\tage)$ depends upon the IMF.

In Figure~\ref{fig:msurv_imf}, we show the general behavior of $\fsurv(\tage),$ including its IMF-dependence, which we have calculated using the python-fsps wrapper \citep{python_fsps} of the FSPS library \citep{conroy_gunn_white_2009_fsps1,conroy_gunn_2010_fsps3}. We show results for four different widely-used IMFs: Salpeter \citep{salpeter_1955}, Kroupa \citep{kroupa_etal_1993_imf}, Chabrier \citep{chabrier_2003_imf}, and van Dokkum \citep{van_dokkum_2008_imf}. Each colored curve is accompanied by a dashed curve providing a parametric fit implemented in DSPS; the bottom panel of Figure~\ref{fig:msurv_imf} shows the difference between the fitting function and the result calculated by FSPS; the functional form we use to approximate $\fsurv(\tage)$ is defined in terms of $\terf(x),$ with parameters specified in the \href{https://github.com/ArgonneCPAC/dsps/blob/main/dsps/surviving_mstar.py}{surviving\_mstar} module of the DSPS package.

Most stars leave behind a remnant when they die, which is either a white dwarf, a neutron star, or a black hole; the physical nature of the remnant depends on the initial mass of the star. In Figure~\ref{fig:mrem_imf}, we show $\frem(\tage),$ the fraction of stellar mass locked up in remnants, plotted as a function of time for the same four IMFs shown in Figure~\ref{fig:msurv_imf}. Each solid each curve in Figure~\ref{fig:mrem_imf} is accompanied by a dashed curve illustrating a triweight-based parameterized fitting function for $\frem(\tage)$ defined in the \href{https://github.com/ArgonneCPAC/dsps/blob/main/dsps/remnant_mass.py}{remnant\_mass} module of the DSPS package.

\renewcommand{\thefigure}{B\arabic{figure}}
\section{Triweight Kernel}
\label{sec:tdubs}

\begin{figure}
\includegraphics[width=8cm]{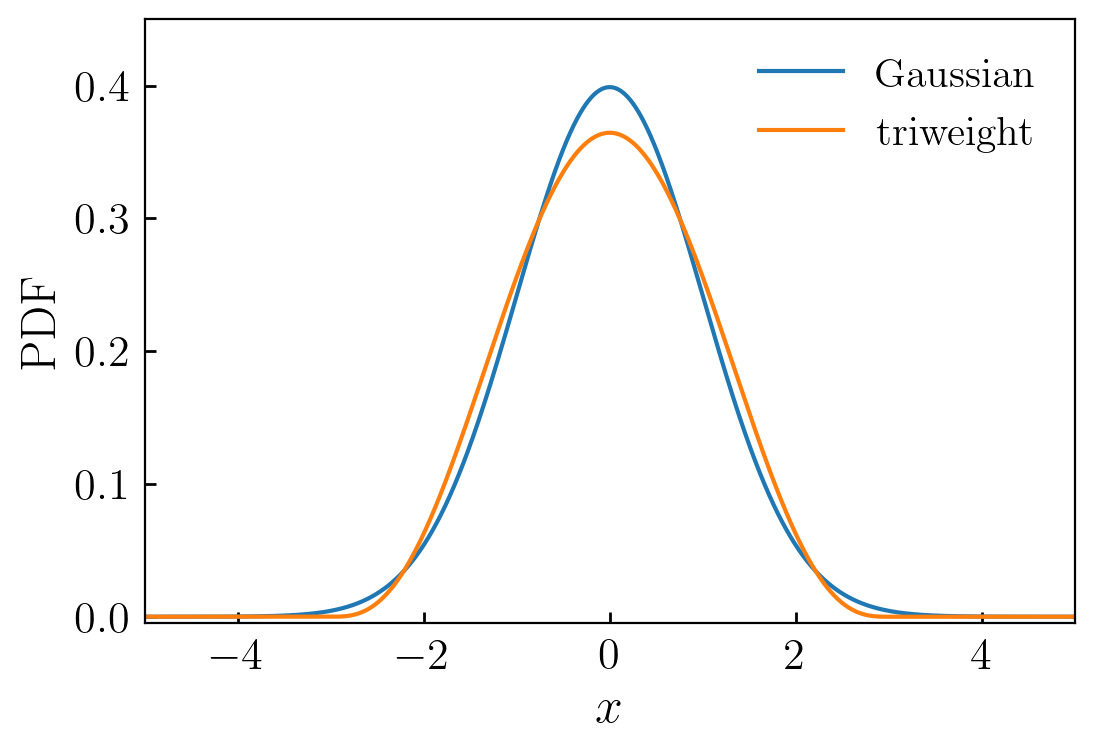}
\caption{{\bf Triweight kernel}. The blue curve shows the probability density function of a normal distribution centered at $\mu=0$ with second moment $\sigma=1.$ The orange curve shows the triweight kernel, $\mathcal{T},$ defined in Eq.~\ref{eq:triweight}, again with $\mu=0$ and $\sigma=1$. The triweight function $\mathcal{T}(x)$ vanishes at points beyond $\vert x-\mu\vert>3\sigma,$ is continuously differentiable for all $x,$ and is highly performant on GPUs due to its formulation in terms of elementary arithmetical operations.}
\label{fig:triweight}
\end{figure}

Many of the calculations in this paper involve probability-weighted summations defined by a clipped Gaussian PDF. In this section, we review an alternative distribution based on the triweight kernel, $\mathcal{T},$ defined as:
\beq
\label{eq:triweight}
\mathcal{T}(x\vert\mu,\sigma) \equiv\begin{cases}
	\frac{35}{96}\left[ 1-(z/3)^{2} \right]^{3}, & \vert z\vert\leq3 \\
	0, & \text{otherwise}
\end{cases}
\eeq where $z\equiv(x-\mu)/\sigma.$
The comparison to a Gaussian is shown in Figure~\ref{fig:triweight}. The two distributions have the same first and second moments, $\mu$ and $\sigma,$ but differ in their higher-order moments. The coefficients appearing in Eq.~\ref{eq:triweight} are defined so that the function $\mathcal{T}(x)$ is $C^{\infty}$ across the real line; points with $\vert x-\mu\vert>3\sigma$ contribute formally zero weight, and so using a triweight is essentially equivalent to using a clipped Gaussian. However, there is a significant computational advantage to the triweight: evaluating a Gaussian requires a special function evaluation that that can be far slower on a GPU accelerator device in comparison to the small number of elementary arithmetical operations required to evaluate Eq.~\ref{eq:triweight}.

Some calculations in this paper are formulated in terms of $\terf(x),$ the cumulative integral of Eq.~\ref{eq:triweight}, such as the $\tage$-dependent metallicity distribution function shown in \S\ref{subsubsec:agedep_metals}. The $\terf$ function is also defined by a simple algebraic expression that is also performant on GPUs:
\begin{gather}
\label{eq:terf}
\mathcal{T}_{\rm erf}(x\vert\mu,\sigma) \equiv\begin{cases}
	p(z) & \vert z\vert\leq3 \\
	0, & \text{otherwise}
\end{cases}\\
p(z) = \frac{1}{2} + \frac{35}{96}z - \frac{35}{864}z^3 + \frac{7}{2592}z^5 - \frac{5}{69984}z^7\nonumber
\end{gather}

\bsp	
\label{lastpage}
\end{document}